# An Approach to Programming Based on Concepts

## Alexandr Savinov

Institute of Mathematics and Computer Science, Academy of Sciences of Moldova
str. Academiei 5, 2028 Chisinau, Moldova
Department of Computer Science III, University of Bonn
Römerstr. 164, 53117 Bonn, Germany

http://www.conceptoriented.com/savinov

**Abstract.** In this paper we describe a new approach to programming which generalizes object-oriented programming. It is based on using a new programming construct, called concept, which generalizes classes. Concept is defined as a pair of two classes: one reference class and one object class. Each concept has a parent concept which is specified using inclusion relation generalizing inheritance. We describe several important mechanisms such as reference resolution, context stack, dual methods and life-cycle management, inheritance and polymorphism. This approach to programming is positioned as a new programming paradigm and therefore we formulate its main principles and rules.

# 1    Introduction

The problem of object identity has been paid significant attention in computer science. In particular, it is considered a pillar of object orientation [Ken91] and one of the corner stones of data modelling [Wie95]. References, keys, surrogates, pointers, names and other identification techniques have been used for many decades. Yet, identity modelling is still much less developed domain when compared with entity/object modelling. In particular, two currently dominating disciplines – object-oriented programming and relation model – are targeted mainly at describing what happens in entities and provide very limited facilities for identity modelling. So far entities/objects have always been first class citizens when compared with identities/references. Although it is widely recognized that references provide very important services, they have always remained somewhere underground having the status of second class citizens. There were many attempts to study identification techniques from philosophical and conceptual points of view, and numerous rather specific techniques were proposed for solving identity-related problems. However, the problem of identity modelling has still secondary importance with respect to entity modelling. Indeed, objects possess structure and behaviour while references do not. Objects are armed with such powerful mechanisms as encapsulation, inheritance and polymorphism while references are deprived of all of these. Objects are at the focus of most system development approaches while references always remain in their shadow. Although references could be modelled manually on the basis of standard facilities, doing so is tricky, difficult and error prone because it is done by means of objects, i.e., by applying inappropriate means. The approach described in this paper aims to fill this asymmetry and disparity between objects and references by making references first class citizens with the same rights as objects.

We discuss the problem of identity modelling in a narrower context of programming languages. In OOP it is assumed that references are primitive elements provided by the compiler taking into account services of the target run time environment. Objects in OOP are in the centre of the program design and therefore class – the main programming construct – is intended to model only objects. References play an auxiliary role of something that is necessary and cannot be completely avoided but what we do not want to see explicitly in our design. For example, let us assume that a variable stores a reference to a bank account object. This variable is then used to access this account balance using some its method. In OOP we are completely unaware of the reference format and the operations executed during object access. The compiler provides primitive (native) references and the programmer has an illusion of instant access to objects. In this case we can model only structure and behaviour of objects such as bank accounts while structure and behaviour of references is completely out of the scope of OOP.

The main problem with this conventional approach is that frequently we need to define our own *custom references* with their own format and associated access procedures. In this case using universal standard references with built-in functionality might be a limiting factor. For example, creating and deleting many tiny objects is known to be a very inefficient procedure. Dealing with remote objects requires special references encoding information on their location and special access procedures based on some network protocol. Using persistent objects is also based on specific requirements to their identifiers and access rules. In all these cases it would be natural to develop a special subsystem which can take into account specific features of its objects. However, an important observation here is that such a subsystem is aimed at modelling how objects are represented and accessed rather than what are object's structure and behaviour. In other words, we see that there exists functionality which is not associated with objects and hence it should be modelled somehow differently. Here again this problem can be solved using special features of operating system (like local heap), middleware (like CORBA or RMI), libraries (like Hibernate) or programming patterns (like proxy). However, our goal consists in developing mechanisms for reference modelling which would be as powerful as those for object modelling. By applying these mechanisms we could model arbitrary references with any format and functions. Just as object classes are developed for each concrete program depending on its design requirements, references also should be modelled for each concrete program reflecting its design goals. For example, we know that in the problem domain bank accounts are identified by their numbers. Therefore, ideally account objects in the program should be also represented by numbers. Yet, OOP forces us to use primitive references for any class of objects. Further, bank account objects might be known to be stored in a database where they can be accessed using some query language. However, we are not able to reflect this directly in our program code because in OOP all objects are accessed in one and the same way.

In this paper we propose a solution which is based on extending an object-oriented programming language by introducing new language constructs and mechanisms. The programmer then does not depend on the available environment and compiler with their standard object identification facilities. Rather, using the proposed approach it is possible to create internal custom containers with a virtual address system where all the objects will exist. In particular, we can develop references of arbitrary complexity including structure and functions. For example, references could contain account numbers instead of native references while objects could really reside in a database on a remote computer or anywhere in the world. At the same time we can still think of objects as being represented and accessed directly. For example, even if an account is represented by its number and resides in a database, we can get its current balance by applying one method. The underlying environment developed by the programmer will intercept this call and perform all the necessary intermediate actions transparently. Getting an account balance might mean using some special protocol involving also operations with some special database developed for this and only this bank. Thus references are interpreted as *virtual addresses* providing at the same time a possibility to bind them to real locations.

Unfortunately, it turned out that the problem is not so simple. Although some mechanisms could be implemented using traditional existing techniques, such partial solutions make the whole picture more complex. One of the main problems is that functions of references and functions of objects have a *cross-cutting nature* and cannot be easily separated or mechanically combined. The structure of references and objects, and relationships between them are discussed in Section 2. In particular, in this section we postulate that any element has two sides or flavours, called identity and entity (Section 2.1), and there are two fundamental relations among elements, called *inclusion* (Section 2.2) and *substitution* (Section 2.3). Both these relations are of crucial importance for understanding the proposed approach. Inclusion generalizes inheritance while substation is a completely new notion.

Further in Section 3 we propose an approach to modelling elements as they are described in Section 2. This approach is based on using a new programming construct, called *concept*. Accordingly, the whole approach is referred to as concept-oriented programming (CoP). Concept consists of two classes, called reference class and objects class (Section 3.1). Instances of the reference class, called references, are passed-by-value and are intended to represent objects, which are instances of the object class passed-by-reference. Definition and properties of this construct are described in Section 3.2 while Section 3.3 discusses how two constituents of one concept are connected using the mechanism of *meta-transition* implemented via substitution relation.

Section 4 is devoted to discussing inclusion relation which generalizes inheritance and hence is of high importance for understanding distinguishing features of the described approach. In particular, this relation is used for defining so called *complex references* (Section 4.1) which are used to identify objects in a hierarchical space. The inclusion structure of elements in the program determines how they are accessed and this standard sequence of access is described in Section 4.2. In order to increase efficiency of access we propose a mechanism, called *context stack*, which is discussed in Section 4.3. In Section 4.4 we describe how elements (objects and references) are created and deleted.

Since references are in the focus of our research, in Section 5 we propose a number of operations which make it easier to manipulate them in the program. In particular, Section 5.2 describes in details reference structure and parameters. Section 5.2 defines operations of left and right casting for references. And Section 5.3 discusses the mechanism of reference length control.

An amazing feature of the proposed approach to programming is that it is backward compatible with OOP. Section 6 describes main differences of these two approaches discussing the mechanisms of inheritance (Section 6.1), polymorphism (Section 6.2) and life-cycle management (Section 6.3).

An approach to programming described in this paper is not simply a technique but rather it turns into a programming paradigm. Why it is so and new principles of programming are described in Section 7. Section 8 is devoted to discussing related work and finally in Section 9 we make concluding remarks.

## 2 Representation and Access

### 2.1 Entities and Identities

We assume that a system consists of *elements* or *things*. One element consists of two parts according to the following concept-oriented principle:

> Principle 1 [Duality]. Any element exists in two forms or flavours called *entity* and *identity*.

Thus elements or things exist in connected pairs rather than as independent parts. Formally, it is not possible to have one part without the other – they always co-exist. Identity is a unique identifier for a location at which the corresponding entity can be found. Thus identity is precisely what we normally mean by address, pointer, reference or name. Entity is what is characterized by some location and it is precisely what we normally mean by object or record.

The main role of identity (but not the only one) is that it serves as a *representative* of its entity and provides *access* to it. If we need to access an entity, pass or store information to it then the only way consists in using its identity. Entities can be manipulated only through their identities, i.e., all operations are performed exclusively with identities. Such a separation of roles corresponds to Kant's view expressed originally in his work "Critique of Pure Reason". According to this view entity is a thing in itself or reality which is not observable in its original form and is radically unknowable. In contrast to the reality it represents, identity is a phenomenon observable directly in its original form. Thus the only way to get, pass or store information about any entity consists in using its identity which always stays between us and the entity.

On the other hand, entities have one property that in great extent justifies their usage – they exist objectively without the necessity to have anything else. In other words, entities are precisely what we normally mean by objective reality which exists independently of any subject or observer. This reality consisting of entities is characterized by properties which can change in time. By getting or setting these properties we can read or write information which will then exist independently of us. So entity can be thought of as a *persistent* part of an element in contrast to identity which is thought of as a *transient* part. If we change an identity then this will be visible only for us and this change disappears along with us. If we modify an entity then this will be visible for anybody who has access to it by possessing its identity. Thus entities are persistent but cannot be directly accessed and therefore we use identities which are directly observable but do not retain their state outside the current scope. This difference is analogous to information stored in local variables and information stored in memory which however can be accessed only using some identities in local variables on the stack.

If entities are not directly observable then how do we know whether some entity exists or not? The only way consists in using identities. This means that if we have an identity then we assume that the corresponding entity really exists (even if it does not). And if the identity is not available and cannot be restored then the corresponding entity is supposed to be non-existing (even if it really exists). Thus identities serve to represent the fact that the corresponding entity exists, i.e., they represent the fact of their existence. There is no other way to check if an entity exists except for getting somehow its identity.

Identities and entities are defined in pairs but on the other hand they have to exist separately. In other words, identities and entities exist in two different worlds but are connected in pairs (Fig. 1). One of the most interesting properties is that we actually do not know how this association between the two parts of each pair is implemented. We manipulate identities which somehow guarantee that there will be some influence on the corresponding entities but how concretely an identity accesses its entity remains a magic. In other words, we do not know how the border between the world of identities and the world of entities is intersected. The moment when a process moves from one world to the dual one is referred to as *meta-transition* (we called it also turn-point in previous works).

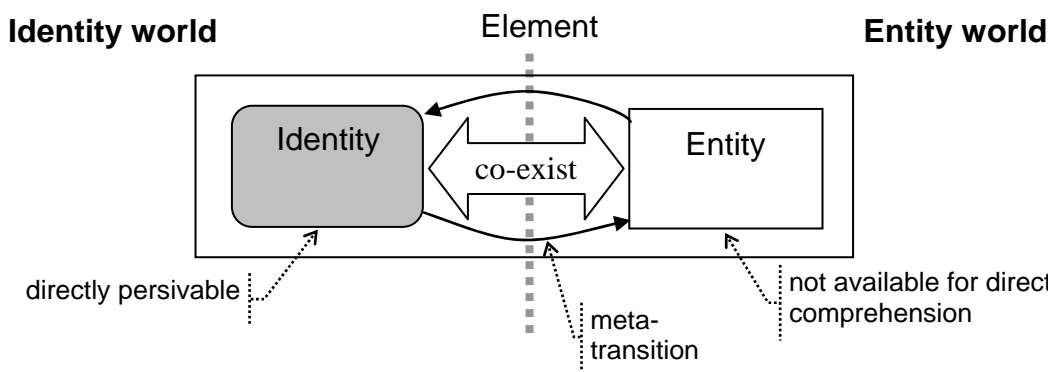

Figure 1. Identity and entity are two sides of one element.

For example, we can store an address of some organization (identity) but the organization itself (entity) is a separate thing. Moreover, we cannot even precisely define what this organization is, except that it provides some services accessible through the available address. It can be a virtual organization or it may have real location in some building with human workers. For us it is important only that having an address we can access its services. Although this address is thought of as the organization location it is not necessary that this entity has a real location. Identity is a *virtual* address and hence this location is simply a convention used to distinguish entities.

Another example of identity is a memory address in computer. Notice however that there may be several types of memory addresses like physical address and virtual address. In any case having such an address is enough to get some content which is thought of as residing at this location. In particular, the address provides a number of services and guarantees that if we store some content at this location then later we can retrieve this same value. An interesting thing is that the real location needs not to be a memory at all or at least we do not need to know where and how the entity (content) is stored. We manipulate only the memory address (identity) and rely on its intrinsic (and magical) capability to interact with the represented entity. Another very important property is that the memory content does not actually store information on its identity (its address in memory). In other words, nowhere in memory is stored that this cell has some address. It is some other mechanism that interprets this cell as having an address. In particular, there is no way for the programmer to change the memory cell address by changing the content of the identity or entity.

In order to illustrate what is not an identity let us consider the mechanism of primary keys in the relational model of data. On one hand a primary key looks like an identity because we can use the selected columns to find the corresponding row. However, strictly speaking it is not so because the primary key is not separated from the row (entity) it belongs to. Thus primary key exists in the world of entities; it is part of an entity with special role. We simply declare that some part of the entity (primary key columns) will be used to store information on it which can be then used to find it. In contrast to identities, primary keys can be modified (just as other columns of the row) and hence we can change the location of the entity. As a consequence it is even not guaranteed that this entity can be found. Compare it with references in OOP or memory addresses which are not stored in the represented entity and cannot be changed as a property of this entity. Indeed, an object class does not have a field with the reference value and objects do not store their own references. Nevertheless it is guaranteed that this object reference will be able to represent it. Thus primary keys are not identities but can be viewed as an attempt to create identification means within the relation model. In this sense primary keys can be viewed as identifying properties. For identity it is important to be detached from its entity and exist separately from it.

## 2.2 Inclusion Relation

In the previous section we postulated that an element is a pair of one identity and one entity. Identities and entities exist separately in their own worlds and identities are used to represent and access entities. One identity is interpreted as a location or address of the entity. However, normally we manipulate *many* elements within some range of their possible identities. This set of similar elements is thought of as a *space*. For example, a set of possible memory addresses is a space. Identity of this element is one address and its entity is the content at this address. A set of all IP addresses is a space where one element is an IP address as identity and one computer as an entity.

Such a grouping of similar elements in spaces is not simply a convenience method. We assume that elements cannot exist without a space, i.e., any element needs some space to exist. In other words, if we have an element (identity-entity pair) then it is necessary to specify some space it belongs to. Such an indication of the space for an element is important because it allows us to compare different elements. For example, a memory address cannot be compared with a postal address because they belong to different spaces. The relation between elements and the space is called *inclusion*, i.e., each element is included in some space and a space can include internal elements.

Having spaces is convenient and important but the problem is that it is not clear how to interpret them. In order to overcome this problem we assume that spaces are normal elements consisting of one identity and one entity (Fig. 2). Thus we still can consider only elements (identity-entity pairs) by adding only inclusion relation among them. This is formulated as the next concept-oriented principle:

Principle 2 [Inclusion]. Any element is *included* in some other element.

Parent elements are thought of as spaces for their child elements.

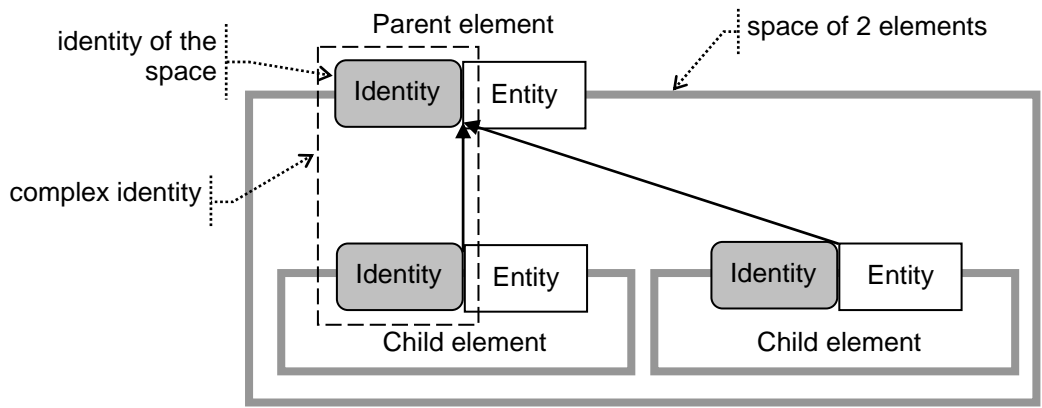

Figure 2. Inclusion of elements in hierarchical space.

It should be noted that by inclusion in this paper we mean *physical inclusion* which has to be distinguished from *logical inclusion* considered in the concept-oriented data model (CoM) [Sav05a, Sav05b, Sav06, Sav07a]. One important characteristic feature of physical inclusion is that child elements do not store any information on their parents such a reference or pointer. We say that a child simply exists in its parent as in some context. Having a parent is an intrinsic fundamental property of any element postulated in the above formulated principle rather than a property that can be assigned or changed. Elements are born within some parent context and then exist there for the whole life-time. They know their parent and can access it but they do not store any reference to it as a property. (For comparison, having a logical parent in CoM means referencing some element.) For example, block on disk does not store information that it belongs to some disk – it just exists in it as in its physical space. A record does not have any indication that it belongs to some table – it simply exists within this table. A star does not have any information on the galaxy it belongs to and so on.

Just as any other element, a space has two constituents: an identity and an entity. The only difference is that spaces may have a number of child elements. If a space has no children then it is considered an ordinary element. For example, a city element has two constituents: city name (identity) and city itself (entity). Street and house also consist of two parts (identity and entity) which need to be always distinguished, i.e., street name is different from street object.

We assume that no cycles exist in the inclusion relation structure which has the form of a tree with one root. For example, postal addresses start from one root which includes countries (say, country code as identity and country properties as entity) which in turn include cities and so on. The structure of the hierarchical space can be variable. For example, domain name system (DNS) consists of a sequence of computer names separated by dots where each next name identifies an element in the context of the previous one. For example, www.conceptoriented.com consists of three elements identified by "com", "conceptoriented" and "www".

We said in the previous section that the main role of identity consists in representing entities and providing access to them. The role of inclusion relation consists in structuring the elements. This structure can be then used to easier identify the entities. In other words, instead of one big space of elements we get a number of smaller sub-spaces and the most interesting feature is that the spaces themselves are normal elements. One advantage of having such a hierarchical space is that local identities within one space can be smaller because they need less space to identify a smaller number of elements. For example, it is not necessary to specify a country if we send a letter within this same country.

However, a disadvantage is that identities from different spaces may be equal and the representation is then not unique. In other words, one and the same identity is assigned to different entities and the only difference is that these elements exist in different spaces. For example, one street name can exist in different cities, one city name can exist in different countries and one memory address can exist on different computers. In order to distinguish such elements the parent identity has to be attached to this element identity. However, the same parent identity may also exist in different spaces and so on. Thus the full representation of an entity consists of all its parent identities and each element is context-dependent. Such a sequence of identities where each next identity is included into the previous one is referred to as a *complex identity*. The operation of attaching a child identity to its parent identity is called *concatenation* or *extension*. One constituent in a complex reference is referred to as an *identity segment* while the corresponding entity is referred to as an *entity segment*. Thus using complex identities in the hierarchical space we can unambiguously represent all the entities. For example, a city is fully represented by specifying its country name as the first segment and city name as the second segment.

## 2.3 Substitution Relation

In the previous section we described a phenomenon of hierarchical space where elements exist within other elements and such an inclusion is postulated as their intrinsic property. Element identity can be used to represent and access the corresponding entity in the context of the parent element. Thus it is important that any identity has a meaning (can be meaningfully interpreted) only in the context of its parent element and cannot be considered outside its context. In other words, we need to know the parent element in order to understand what this element means. In this section we focus on a different phenomenon which also allows us to determine the meaning of an element but in the other way. Both mechanisms then represent two sides of one thing and should be considered together.

Let us consider a simple address space consisting of all computer names (without a hierarchy for simplicity). One name is an identity representing one entity – a computer. Using this name we can access this computer but we do not know how the connection from the identity to the entity is implemented. This addressing schema is simple and clear except for one subtle observation: we know that in addition to name each computer has also some IP address assigned to it which also can be used as its identity. Thus we have two identities for one and the same entity. We can represent and access any computer using its name or we can choose to access it using its IP address. The same situation exists in programming where one and the same object is represented by its primary key, by its Java reference, by its remote reference, by its memory handle or by its physical address.

The second important observation is that different identities of one and the same entity are not arbitrary and are somehow connected. For example, computer name is connected with its IP address and Java reference of an object is connected with its memory handle which in turn is connected with the object physical memory address. This relation is referred to as *substitution* and the procedure for finding the substituted identity is referred to as *resolution* (Fig. 3). (Their dual analogues are inclusion and extension described in the previous section.) Thus we say that computer name substitutes for an IP address, a Java reference substitutes for some memory handle and a remote reference substitutes for a Java reference. Thus one and the same entity may have many identities which are connected by the substitution relation. We generalize these observations in the following concept-oriented principle:

Principle 3 [Substitution] An element *substitutes* for some other element.

Notice that substitution relation describes both identities and entities, i.e., an entity also has a substituted entity.

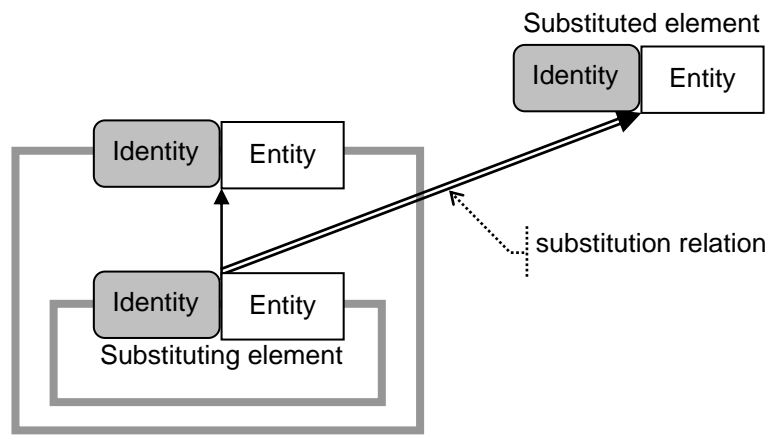

Figure 3. Substitution of elements.

We will assume that substitution is not independent but rather is constrained by inclusion relation. In particular, we assume that an element cannot substitute for its child elements. More precisely, an element can substitute for only siblings of some of its parent. This property will be formulated more precisely later when types of elements will be introduced. Now it is important to understand that the role of substitution consists in providing a way to basic elements. For example, remote references are resolved to Java references which in turn are resolved to memory handles and so on. In this sense substitution is opposite to inclusion.

The above formulated principle allows us to partially answer the question about the transition from the identity world to the entity world. Earlier we said that it is a magic and we do not know how identities get access to the entity world. It is still so but now we can reduce this procedure to more basic things. Namely, we say that an identity performs meta-transition into the entity world using the substituted identity. This means that it is the root identity that is responsible for all the interactions with the entity world. All other identities directly or indirectly substitute for the root identities and hence they are resolved into them when it is necessary to perform real access. How meta-transition is performed by the root identity is not known – it is considered its intrinsic or built-in ability.

For example, if we have a computer name and want to get some information from it then the access procedure will resolve this name to the substituted IP address which will be then used for access. If we use Java reference for identifying our objects then each access will result in the reference resolution. This procedure will restore the object memory handle which will be used for access.

Theoretically the resolution proceeds infinitely deeply by restoring more and more basic identities which are closer to the reality. Yet no one of them provides direct access to the reality (to the entity world). For example, memory handles are resolved into physical addresses which are then processed by the CPU which then communicates with the memory chip and so on. In practice, we choose some level as a basic one and do not go deeper. This level can be viewed as a base coordinate system. All other spaces and identities provide indirect representations for its coordinates. For example, we can choose memory addresses as a base coordinate system and then introduce indirect coordinates in the form of Java references. Or we can choose Java references as a base coordinate system and then introduce remote references for their substitution.

Substitution relation allows us to introduce virtual identifiers which are not directly bound to the base coordinates. Such virtual identifiers bring a new level of indirection and hence flexibility. In particular, having a virtual identity of an element we can control its real coordinate in the base space. For example, we can change IP address of a computer and this allows us to reuse its name stored in different places.

# 3 Concepts for Describing Entities and Identities

## 3.1 Reference Class and Object Class

In the previous section we assumed that elements consisting of one identity and one entity are organized into a hierarchy where each element has one parent and one substituted element. The question is then how can we describe structure and functions of identities and entities? One wide spread approach in computer programming consists in using classes for that purpose. Class fields allow us to describe structure while class methods describe its behaviour. What is new in our approach is that we need two sorts of classes for describing two parts of an element. In other words, we need one class for describing structure and functions of identities and one class for describing that of entities. Since further in the paper we will focus mostly on computer programming issues, we change our terms to more specific. Namely, instead of the term identity we will use the term *reference* and instead of the term entity we will use the term *object*. Thus our next assumption is that an element of a computer program consists of one reference and one object which are modelled by one *reference class* and one *object class*, respectively.

Object classes are equivalent to normal classes as they are defined and used in OOP. Instances of the object class are referred to as *objects*. Thus if no reference classes are defined then we get an object-oriented program which consists of objects, i.e., its structure and functionality is concentrated in objects. Reference classes are used to describe object representatives and their instances are referred to as *references*. In this case a program consists of two types of elements – objects and references – and the program functionality is distributed among them. Thus references are as important as objects and in some cases can even account for most of the program complexity.

For example, let us assume that we need to model bank accounts. An account consists of its identity and its entity which are modelled by two classes. If an account is identified by its number having balance as a state then it can be described using two classes as shown in Listing 1.

```
01  reference AccountReference { // Reference class
02     String accNo; // Identifying field
03     ... // Other members of the reference class
04  }
05
06  object AccountObject { // Object class
07     double balance; // State field
08     ... // Other members of the object class
09  }
```

Here we use keyword 'reference' to mark a class as a reference class and keyword 'object' to mark a class as an object class. We might also add other members to these classes, say, opening date field or a method for getting balance. These two classes have their own names and we can produce their individual instances in the source code. The main goal of introducing reference classes consists in using custom references for representing objects rather than only primitive references. For example, if we want to create a new account object having custom identifier then we can specify two class names. Using the style of the Transframe programming language [Sha] this can be written as follows:

```
AccountObject account of AccountReference;
```

This declaration means that variable `account` will contain reference of class `AccountReference` which represents an object of class `AccountObject`. Using the style of C++ smart pointers [Str91] the same can be written as follows:

```
AccountReference<AccountObject> account(new AccountObject);
```

Here again we declare a variable which contains a reference of one class pointing to an object of another class. In any case it is important that the two classes are defined independently and we need to specify both of them when declaring new variables, parameters or return values. In particular, one reference type can be used for representing different objects and one object can be represented by many different reference types.

## 3.2 Concept Definition

Using separately reference classes and object classes is possible but is not very convenient. In this case references and objects are defined separately and there is no connection between them because these classes are paired only when they are used for declaring a variable. Thus the classes describing one element cannot use each other.

To overcome this problem we propose to use a new programming construct, called *concept*, which combines two classes: one reference class for describing identity and one object class for describing entity. The reference class and the object class loose their independence and can exist only as part of one concept. If we need to have only a reference class then it can be defined as a concept with the empty object class. And if we need to have only an object class then it can be defined as a concept with the empty reference class. The latter is equivalent to normal classes.

```
01  concept Account // One name for the pair of two classes
02     reference { // Reference class of the concept
03        String accNo;
04        ... // Other members of the reference class
05     }
06     object { // Object class of the concept
07        double balance;
08        ... // Other members of the object class
09     }
```

For example, instead of defining separately one account class and one account reference class (Listing 1) we now need to define one concept which contains the both (Listing 2). It will be assumed that concepts start from the keyword 'concept' followed by the concept name. Then the reference class is defined using keyword 'reference' and the object class is defined using keyword 'object'. The two classes are described as usual by defining their members.

Since reference classes and object classes cannot be defined independently, they cannot be used separately in the program. Instead, we need to use concepts only, i.e., named pairs of one reference class and one object class. An approach where concepts are used instead of classes for declaring types is referred to as *concept-oriented programming* (CoP). Thus concepts in CoP are used where classes are used in OOP when declaring a type of variables, fields, parameters, return values etc. For example, let us consider the following code:

```
Account account = getAccount();
Person person = account.getOwner();
Address address = person.getAddress();
```

Strictly speaking, it is not possible to determine if it is an OOP program or CoP program from this fragment because we do not know how its types (`Account`, `Person`, `Address`) are defined. If they are defined as normal classes then it is an OOP program. Otherwise, if they are defined as concepts (for example, as concept `Account` in Listing 2) then it is a CoP program.

So what actually changes when we use concepts instead of classes? The main change is that variables declared using concepts contain custom references in the format defined by the concept reference class. In contrast, if a variable is defined using a class then it contains a primitive reference chosen by the compiler. Thus changing concept definition we can effectively influence what is actually stored in the variables in the program. The main goal of such custom references consists in controlling how objects are represented and accessed (although there are other uses).

For example, if `Account` is a concept defined in Listing 2 then variable `account` will store account number as its content. This account number is supposed to indirectly represent some account object and it is passed to method parameters, stored in object fields or returned from methods as if it were normal (primitive) object reference. Analogically, if `Person` is a concept then variable `person` would store passport number and birth date which both indirectly represent the owner of the account.

Listing 3. Precedence of reference methods.

```
01  concept Account
02    reference {
03      String accNo;
04      double getBalance() { // Reference method
05        print("=== Account::getBalance reference method");
06        return 0;
07      }
08    }
09    object {
10      double balance;
11      double getBalance() { // Object method
12        print("--- Account::getBalance object method");
13        return balance;
14      }
15    }
16
17  Account account = getAccount();
18  double balance = account.getBalance();
19
20  $ === Account::getBalance reference method
```

Two classes constituting a concept can provide their own members. In particular, the reference class and the object class can define one and the same method (a method with the same signature). Such methods defined in both classes are referred to as *dual methods* and play very important role in CoP. The definition provided in the reference class is called a *reference method* while the definition provided in the object class is called an *object method*. (Theoretically it is more convenient to assume that both definitions are always available and if one of them is absent then some default implementation exists.) Yet in the source code we use methods as usual with no indication what kind of method has to be used. For example, if method `getBalance` is defined in both the reference class and the object class of concept `Account` and we apply it to variable `account` of this concept (Listing 3, line 18) then what definition has to be actually executed?

In order to resolve this ambiguity we use the following principle:

> Principle 4 [Dual methods]. Reference methods of a concept have precedence over its object methods.

In other words, applying a method to a variable means executing the reference method rather than the object method. For example, the statement `account.getBalance()` will use the definition provided in the reference class of concept `Account` (Listing 3). This means that in the source code we can act only on what is stored in variables by value, i.e., we work with references but assume that they should affect the represented objects. By calling method `getBalance` we want to access the account object however this operation cannot be executed directly and the reference stored in the variable intercepts this call and executes all the necessary intermediate actions.

This principle means that references intercept any access to the represented object. It is quite natural because the object method cannot be executed because of the absence of its primitive reference. Indeed, we cannot apply method `getBalance` to simply an account number stored as a string and hence some intervention is needed. The reference is then the only element that knows where the object is located and how to access it, particularly, how to call its methods. References protect objects from direct access from outside and there is no way to bypass a reference if it wants to control access to the represented object. Notice that the programmer has still the illusion of working with the object directly because the interception is performed transparently. If an element is associated with a space border then any access request is processed in the reference on one side and only after that it can cross the border (Fig. 4). In CoP, methods are thought of as processing channels or tunnels where access requests are processed. In contrast, the standard approach is that methods are thought of as complex operations defined via other operations. In CoP, if we specify a method then it means the name of the path this request needs to follow when travelling to the target object and intersecting intermediate borders.

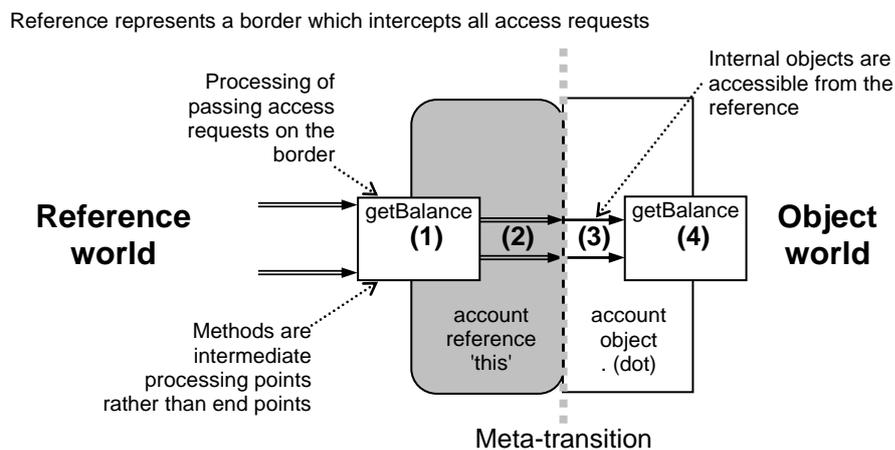

Figure 4. Reference intercepts accesses to the object.

Once a reference method got control it can decide how to proceed. In particular, it may call object methods including the dual object method, i.e., its version defined in the object class. Notice that accessing an object is only possible if we are in its reference. As we already noticed, if a reference method has not been defined then it is simply assumed that there is some (natural) default implementation. Thus from a reference method we can call either this reference methods or this object methods. In order to distinguish them a programming language needs some syntactic means. We will use a simple convention that this reference is denoted by the keyword 'this' while this object is denoted by dot. For example, `this.getBalance()` is a call of the reference method while

`.getBalance()` is a call of the object method (which can be called only from within some reference method).

Listing 4 provides an example where a reference method calls the dual method and the sequence of access is shown in Fig. 4. If such method is applied to an account variable then first the reference intercepts this call and executes the necessary intermediate operations (step 1 in Fig. 4). Then it calls the dual method using dot to indicate that it is an object method (step 2). At this point the process performs meta-transition and moves to the object world (step 3). And finally the object method executes the necessary operations with this object and returns some result (step 4). The output of this method call is shown at the end of Listing 4.

Listing 4. Call of object method.

```
01  concept Account
02    reference {
03      String accNo;
04      double getBalance() {
05        print("=> Account::getBalance reference method");
06        balance = .getBalance(); // Object method is invoked
07        print("<= Account::getBalance reference method");
08        return balance;
09      }
10    }
11    object {
12      double balance;
13      double getBalance() {
14        print("-> Account::getBalance object method");
15        return balance;
16        print("<- Account::getBalance object method");
17      }
18    }
19
20  Account account = getAccount();
21  double balance = account.getBalance();
22
23  $ => Account::getBalance reference method
24  $ -> Account::getBalance object method
25  $ <- Account::getBalance object method
26  $ <= Account::getBalance reference method
```

### 3.3 Meta-Transition

Program objects represented by custom references can be manipulated as if they were normal directly accessible objects, i.e., in CoP we retain the complete illusion of direct instant access on custom references. For example, we assume that even if account objects are represented by their numbers we still can call their methods as if they were represented by primitive references. In particular, we assumed in the previous section that reference methods can access the corresponding object methods. For example, we can call `getBalance` object method from within `getBalance` reference method:

```
reference {
  ...
  double getBalance() { // Reference method
    return .getBalance(); // Object method call
  }
  ...
}
```

However, it is still not clear how can we call object methods if the object itself is not available because its primitive reference is absent. Indeed, all variables representing accounts store only some account number and it may well happen that the object is not in memory at all and its primitive reference does not exist yet. Account numbers are virtual addresses which cannot be directly used for access and need to be somehow converted into a primitive reference. This means that we need some procedure for intersecting the space border separating two worlds, i.e., a procedure for meta-transition (between steps 2 and 3 in Fig. 4).

In order to provide a mechanism for meta-transition we propose to use concept methods which are intended to implement this border intersection logic and are referred to as *special methods*. In other words, only special methods know how to intersect this border described by this concept. There can be different special methods with different purposes but in this section we describe a method which is intended for normal object access and is called a *continuation method*. Thus the continuation method is used whenever it is necessary to cross the border between the two worlds with the purpose of accessing the object. (Other special methods intended for life-cycle management are described in Section 4.4) All other methods of concepts are executed as if the border would be completely transparent as described in the previous section.

Meta-transition starts in the reference and ends in the corresponding object, i.e., on the other side of the border. Thus it consists of two parts: one is defined in the reference class and the other is defined in the object class. Then the question is how concretely meta-transition is implemented in the continuation method? The idea of the solution is that it resolves the current reference by restoring the substituted reference and then uses its continuation method for meta-transition. It is precisely the point in the program execution where the substituted reference comes in play. It is interesting that the continuation method does not actually perform meta-transition but rather reduces this task to the same method of the substituted reference. The resolved reference in turn finds its own substituted reference and recursively calls its continuation method. This process continues till the primitive reference which is now responsible for real meta-transition. However, we never know how it happens because this level is not controlled by the programmer. When the border is intersected the process finds itself in the object world. At this point the dual part of the continuation method starts. This method is intended for preparation of this object for access.

The sequence of access using the dual continuation methods is shown in Fig. 5. Notice that in comparison with Fig. 4 it has two continuation methods around the border and between `getBalance` methods. If we apply a method to a reference then the reference method will be executed (step 1). If it calls some object method or otherwise accesses the object then meta-transition is executed implicitly using the continuation methods. First, the reference continuation method resolves this reference (step 2) into a primitive reference which again is used for continuation. At this moment the border is interested (step 3). Then the object continuation method starts automatically from which then the target object method is called (step 4).

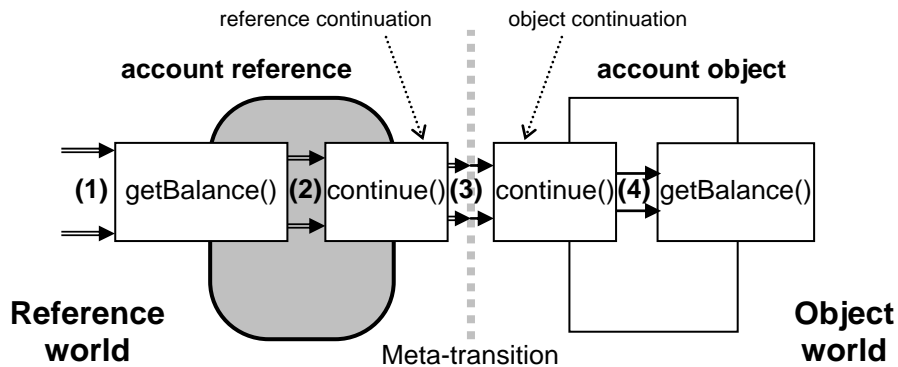

Figure 5. Meta-transition.

Listing 5 gives an example of the continuation methods which demonstrates the sequence of object access and meta-transition. The continuation method is named `continue` and is defined in both the reference class and the object class. The main role of the reference continuation method consists in converting this reference into the substituted primitive reference and then delegating to it the job of meta-transition by recursively calling its continuation method. In this example we assume that the primitive reference is stored in one of the reference class fields (line 4). If this field is not initialized then we assume that the object state is in some secondary storage and needs to be loaded in memory (line 9). For example, here we might load the state of the account from a database using the account number as a primary key. When the substituted primitive reference is available we can really cross the border and go to the object by applying to it recursively the continuation method (line 11). This line is very important because it is rather typical style for the concept-oriented programming when no method is called but the necessary actions are executed implicitly. In this case we simply say that we want to continue and the compiler then implements the logic of meta-transition at the level of primitive references. Thus line 11 is where the object continuation method (line 24) starts. This method is guaranteed to be executed before any access and its main role consists in preparing this object for access. In this example we simply set a flag (line 26) which is a signal for other processes that this object is being accessed. When the object is ready for access we again continue (line 27) but in this case the compiler interprets this method call as a point where the target method called explicitly by the programmer can be executed. If the programmer called `getBalance` then it will start at this point. When the target method or any other access request finishes, the process returns to the object continuation method where the access flag is reset (line 28). After that it returns to the reference continuation method where the state of the object can be stored back into the database if necessary (line 15). And finally the process returns to the reference part of the target method such as `getBalance`. Thus the sequence starts from the explicitly called reference method which wraps implicitly called continuation methods (reference and object continuation) which wrap the object method.

Listing 5. Continuation method.

```
01  concept Account
02    reference {
03      String accNo;
04      Root accObject;
05
06      void continue() {
07        // Resolve account number start access
08        if(accObject == null)
09          accObject = loadAccount(accNo);
10
11        accObject.continue(); // Proceed to the object (3)
12
13        // Clean up and finish access
14        if( lowMemory() ) {
15          saveAccount(accNo, accObject);
16          accObject = null;
17        }
18      }
19      ...
20    }
21    object {
22      double balance;
23      boolean isAccessed;
24      void continue() {
25        // Enter object and prepare it for access
26        isAccessed = true;
27        continue(); // Proceed to the method (4)
28        isAccessed = false;
29        // Clean up and finish access
30      }
31      ...
32    }
```

# 4 Concept Inclusion

## 4.1 Complex References

In the previous section we assumed that one concept describes one space border and discussed how this border between identity and entity is crossed using the continuation method. Then the question is how can we describe a nested structure of spaces and how objects are represented and accessed in this structure?

To model the nested structure of spaces each concept specifies a *parent* concept which describes the parent space. This relation is called an inclusion relation and is denoted by '<' (less than). If concept *B* is included in *A* then it is written as *B*<*A* (*B* is less than *A*). This notation is used in the theory of ordered sets, including formal concept analysis (FCA) and 'less than' sign relates to the number of elements in the sets. Thus the parent concept is supposed to have more elements than its child concepts if they are interpreted as sets. The parent concept is also called a super-concept or base concept while the child concept is called sub-concept or extension concept. The inclusion relation can be also interpreted as a specific-general relation. In this case we say that parent concepts are more general than their child concepts, which are more specific then their parent concepts.

Each concept has some position in the inclusion hierarchy where it has one parent and a number of child concepts. It is assumed that this hierarchy has one root concept with the name *Root*. The root concept describes the outermost space which directly or indirectly contains all elements and is interpreted as the most general concept: $\forall C,\ C < Root$. All events in the program happen within this space and hence it can be viewed as the program itself. Normally this concept is provided by the compiler.

In source code, concept inclusion will be specified using keyword 'in' followed by the parent concept name. For example, if concept `SavingsAccount` is included in concept `Account` then we write it as follows:

```
concept SavingsAccount in Account
   reference { ... }
   object { ... }
```

Informally this means that savings accounts will be located within accounts. It is important that one account may include many savings accounts as well as other more specific account types. If no parent concept is specified then by default it is assumed to be the root concept.

Earlier we assumed that a variable stores a reference in the format defined in its concept. For example, a variable of `Account` concept will contain account number which is a field in this concept reference class. However, in the case a concept has a parent concept then its variables will contain also all the parent references. For example, a variable of concept `SavingsAccount` will contain two references: one defined in `SavingsAccount` and the second defined in the reference class of `Account` concept (Fig. 6). A savings account is then indirectly represented by two numbers: the main account number and the sub-account number identifying its object within the main account.

Inclusion relation allows us to define several new notions. A construct consisting of a sequence of references where each next reference is included in the previous one is referred to as a *complex reference*. Each constituent of the complex reference is referred to as a *reference segment*. A complex reference which starts from the root segment is called a *global reference* (or fully qualified reference). Otherwise, if a reference starts from some internal concept then it is called a *local reference*.

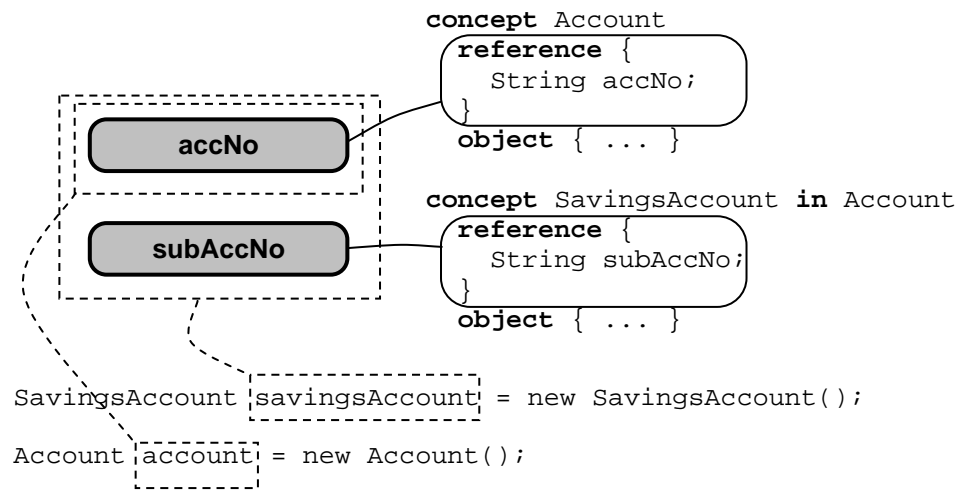

Figure 6. Structure of complex reference.

An object represented by a reference segment is referred to as an *object segment*. Thus the reference segment and the object segment are instances of the reference class and the object class of one concept. A number of objects represented by segments of a complex reference are referred to as a *complex object*.

An important property of complex references is that their segments are stored and passed together as one data structure. It is analogous to how objects in OOP are normally stored in one chunk of memory one segment next to the other (side-by-side). So inclusion relation for reference class can be thought of as a normal inheritance. A difference from objects in OOP is that references need not to start from the root segment because local references may start from any concept. However, by default all variables are supposed to contain fully qualified references with all segments starting from the root and ending with the concept which is a type of this variable. For example, if concept `C` is included in `B` which is included in `A` then a reference to `C` will consist of three segments with the format defined by the reference classes of concepts `A`, `B` and `C` (Fig. 7). These three reference segment represent three object segments which constitute one complex object. However, the three object constituents need not to reside next to each other.

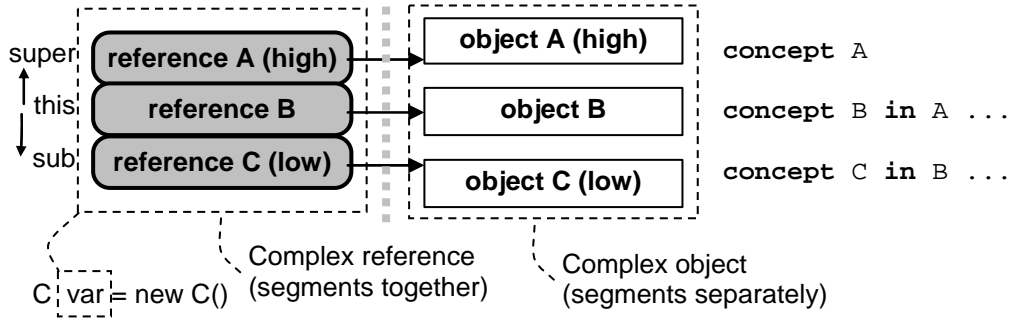

Figure 7. Complex reference and complex object.

Concept instances exist in a hierarchy at run-time and this hierarchy is modelled by concept inclusion relation. Any element has one parent element and a number of child elements. In order to distinguish and access these elements we will use three keywords: 'this' refers to the current element, 'super' refers to the parent element and 'sub' refers to the child element (if any).

## 4.2 Sequence of Access

Earlier we postulated that reference intercepts all accesses to the represented object and object members can be accessed only from within some reference method. However, if a variable contains complex reference then each its segment can implement the same method. For example, both concepts `Account` and `SavingsAccount` can implement method `getBalance` in their reference class. Then the question is which of these two methods has to be executed if this method is applied to an instance of concept `SavingsAccount`? In other words, the problem is to determine which reference segment should process all incoming access requests.

In order to resolve this ambiguity we use the following principle:

> Principle 5 [Reference method overriding]. Parent reference methods have precedence over (override) child reference methods.

In other words, parent reference methods of higher segments intercept all accesses to the child reference methods of lower segments. In our example, `getBalance` of `Account` reference will be called first and only after that it is possible to call `getBalance` of `SavingsAccount` reference.

This principle is quite natural and simply reflects the fact that any attempt to enter a space must be intercepted at the border (Fig. 8, left). Higher segments represent external spaces while lower segments represent internal spaces. In order to reach an element we always start from the external space and then proceed by entering narrower scopes (if we are not already inside). At each intermediate border the request is intercepted by the method having the same signature and the programmer can perform the necessary actions. (Of course, if such a processing is not needed then the interception can be optimized, i.e., if the reference class of the parent concept does not define the method then it will not be intercepted and then the child method can be called directly.) Such a sequence of access effectively means that the only possibility to access an object consists in intersecting all the intermediate borders that separate it from the outside world.

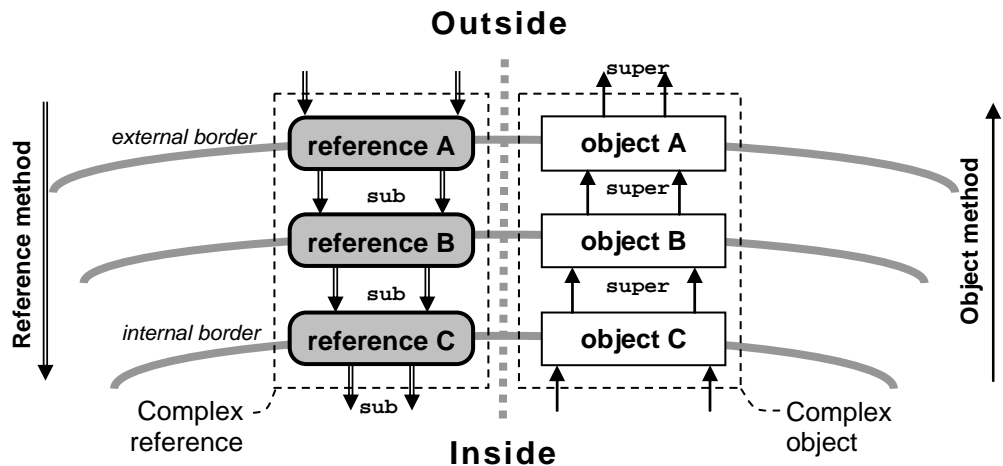

Figure 8. Duality of method overriding.

An important application of this principle is the mechanism of method overriding. However, the direction of such overriding is opposite to the conventional one (as used in OOP), which means that base reference methods override child reference methods. (Notice that this principle belongs to reference methods only.) It is a dual form of the mechanism of method overriding in OOP, which allows a sub-class to provide a more specific implementation of a method defined in its super-class.

Using this principle it is always possible to override any reference method by defining the same method in the reference class of its base concept. Thus parent reference methods protect child reference method from direct use from outside. It is quite natural principle because it allows any border to control incoming processes. For example, a live cell has a border which checks and controls anything that tries to come in. The existence of the own border is actually a generic property of any system including physical, live and social ones. And it is quite natural that if a system is included into another system than the only way to access it from outside consists in intersecting the parent system border. In this sense reference methods can be thought of as incoming methods for the space described by this concept. The mechanism of reference methods and the inverse principle of overriding allow us to support this approach in programming languages. Such a principle of overriding is analogous to that used in the Beta programming language [Kri83, Kri87, Kri89] and the mechanism of inner methods [Gol04].

Listing 6 provides an example illustrating the mechanism of reference method overriding. A variable contains a complex reference of concept `C` which is included in concept `B` which is in turn included in concept `A`. Reference classes of all the tree concepts define method `myMethod` which simply calls the same method of the child reference using keyword 'sub' and prints two diagnostic messages around this statement. (It is assumed that if there is not child then access on 'sub' keyword is equivalent to no-op.) If now we apply this method to the variable of concept `C` then it will print the output shown in lines 37-42. Here we see that any method call is wrapped into a sequence of reference methods starting from the first parent and ending with the last child.

The border between reference and object can be crossed at any moment using the mechanism of meta-transition implemented in the continuation method as described in Section 3.3. Now let us assume that after meta-transition the process has found itself in the object world in some object method. Here we can call object methods or read/write object fields as usual. However, object methods might have been defined in different concepts (in their object classes). In this case there is some ambiguity concerning what definition of the called method to use for execution. This is actually the same question that has been asked in the beginning of this section for reference methods. However, the answer has the dual form. Namely, for object methods we adapt the conventional OOP principle with the normal direction of access (Fig. 8, right):

> Principle 6 [Object method overriding]. Child object methods have precedence over (override) parent object methods.

If we call some object method which is implemented in all object classes in the concept inclusion hierarchy then the compiler will use the definition provided by this object class (we say, that this method overrides its parent methods). After that this method can continue by calling its parent methods using keyword 'super'. Thus child object methods protect parent object methods from direct use from inside. Compare it with the role of reference methods which protect the space from outside. Informally this means that if we need some service or support from an object then the most specific one will be provided first while more general services cannot be accessed directly by internal objects.

Listing 6. Reference method overriding.

```
01  concept A
02    reference {
03      void myMethod() {
04        print("=> A: enter space");
05        sub.myMethod(); // Go inside
06        print("<= A: exit space");
07      }
08      ...
09    }
10    object { ... }
11
12  concept B in A
13    reference {
14      void myMethod() {
15        print("  => B: enter space");
16        sub.myMethod();// Go inside
17        print("  <= B: exit space");
18      }
19      ...
20    }
21    object { ... }
22
23  concept C in B
24    reference {
25      void myMethod() {
26        print("    => C: enter space");
27        sub.myMethod();// Go inside
28        print("    <= C: exit space");
29      }
30      ...
31    }
32    object { ... }
33
34  C myVar = new C();
35  myVar.myMethod();
36
37  $ => A: enter space
38  $    => B: enter space
39  $       => C: enter space
40  $       <= C: exit space
41  $    <= B: exit space
42  $ <= A: exit space
```

An example shown in Listing 7 demonstrates the sequence of access on object methods and the logic of object method overriding. In fact, this program can be produced almost mechanically from the program in Listing 6 by moving method definitions from the reference classes to the object classes and using 'super' instead of 'sub'. We also assume that if no reference method has been defined then its default implementation is to pass control to the child (to more specific element) as shown in the previous example. And if it is the last segment with no child then meta-transition is performed and the object method is called. If method myMethod will be applied to a variable of concept C then it will produce the output shown in lines 37-42.

Listing 7. Object method overriding.

```
01  concept A
02    reference { ... }
03    object {
04      void myMethod() {
05        print("-> A: enter service");
06        super.myMethod(); // Go deeper
07        print("<- A: exit service");
08      }
09      ...
10    }
11
12  concept B in A
13    reference { ... }
14    object {
15      void myMethod() {
16        print("  -> B: enter service");
17        super.myMethod();// Go deeper
18        print("  <- B: exit service");
19      }
20      ...
21    }
22
23  concept C in B
24    reference { ... }
25    object {
26      void myMethod() {
27        print("    -> C: enter service");
28        super.myMethod();// Go deeper
29        print("    <- C: exit service");
30      }
31      ...
32    }
33
34  C myVar = new C();
35  myVar.myMethod();
36
37  $     -> C: enter service
38  $   -> B: enter service
39  $ -> A: enter service
40  $ <- A: exit service
41  $   <- B: exit service
42  $     <- C: exit service
```

If we combine these two examples (Listing 6 and 7) then the following output will be produced:

```
$ => A: enter space
$   => B: enter space
$     => C: enter space
$     -> C: enter service
$   -> B: enter service
$ -> A: enter service
$ <- A: exit service
$   <- B: exit service
$     <- C: exit service
$     <= C: exit space
$   <= B: exit space
$ <= A: exit space
```

According to these two principles any process enters a space via some reference method and then goes down along the inclusion hierarchy to more specific reference segments (Fig. 8, left). Then the process switches to the object world where it changes the direction and goes up to more general object segments (Fig. 8, right). So here it is important to understand that changing the current world (between object world and reference world shown as left and right parts in Fig. 8) entails change of the principle of method overriding and other mechanisms to the dual version. Thus the role and properties of reference methods and object methods are significantly different within inclusion hierarchy. Reference methods are used to enter a scope, i.e., they are executed when an access request needs to come into the space. Object methods are viewed as services intended to be used by internal elements which are already in this space. Actually it is analogous to the natural sequence of access used in real systems and organizations. For example, to get a service from some organization it is necessary to enter its scope and then from inside we can use its internal services.

Diagram shown in Fig. 9 is an alternative view of the generic sequence of access described in this section. Here reference world is shown in the upper part while object world is shown in the lower part. The upper part is what we directly manipulate in the program, for example, by creating a reference of concept `C` consisting of three segments. The three object segments represented by this reference exist in the object world and they can be accessed only using meta-transition (the horizontal line in the middle) implemented in the continuation method (see Section 3.3). Notice that meta-transition can be performed from any reference, i.e., at any intermediate border. However, in this example we assume that the process goes down till the last segment where it switches to the object world and then continues its execution via object methods. The access always proceeds in the downward direction however in the middle it changes its semantics.

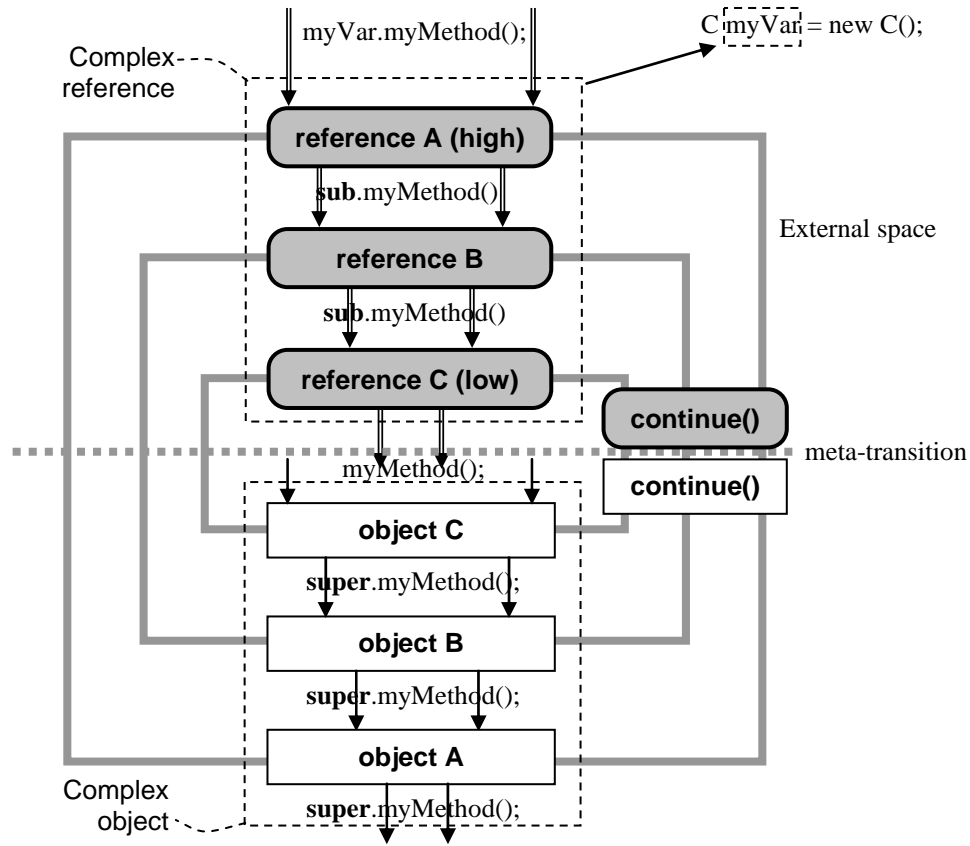

Figure 9. Generic sequence of access.

## 4.3 Context Stack

Let us assume that there is a complex reference which is used to execute some method of the represented object. This target object method as well as intermediate methods executed during access can call methods of the parent objects using keyword 'super' in the same way as it is done in OOP. For example, a method of concept `SavingsAccount` could call methods of its base concept `Account`:

```
concept SavingsAccount in Account
  reference { ... }
  object {
    void myMethod() {
      Person owner = super.getOwner();
      ...
      double limit = super.getCreditLimit();
      ...
      bool isLocked = super.isLocked();
      ...
    }
  }
```

Formally, each access to any object can be performed only via its reference. If we access a parent object then it is represented by a parent segment of the current complex reference and hence this segment has to be resolved for each such method call. In the above example, we would need three resolutions of the parent reference segment for each of these method calls. According to this approach one access requires one reference resolution and after the access is finished the result of the resolution is lost. Although in many cases such as access of parent object in the above example we know that the result of resolution will be the same, the continuation method will still be executed. If there is many parent method calls and field read/write operations then performance of method execution can be rather low because it requires multiple repeated resolutions of the same references even if they are known to be the same.

Such an approach is not only inefficient but is also not natural. It is analogous to the situation where we would need to exit and enter again a building in order to contact several persons in it for different questions. The natural approach would be to enter the building and not to exit before we finish all the necessary actions in it. In other words, in order to execute several operations in some space, its border has to be crossed (resolved) only once and after that the object representing this border and internal objects should be accessible directly without resolution. In the case of a complex reference this means that all its segments have to be resolved in advance and the result of the resolution stored somewhere for future use. After that all the intermediate objects can be accessed directly as many times as needed using the saved primitive references. Thus reference segments are resolved *before* real access takes place and the result of the resolution is stored in a special data structure called *context stack* (Fig. 10). The resolution sequence starts from the first (high) segment and then proceeds to the next segments ending with the last (low) segment. The result of each resolution is pushed on the context stack which grows as each next segment is resolved.

For example, let us assume that an object is represented by a complex reference consisting of three segments of concepts A, B in A, and C in B. Initially, just before the access procedure starts, context stack is empty. When the first segment A is resolved by means of its continuation method it contains a primitive reference to the first object segment of concept A. Since this moment this object is directly accessible. The result of resolving the second segment B is pushed on top of the context stack on the second step, which now contains two primitive references and so on till the last segment. Finally, the number of elements on it is equal to the number of segments in the complex reference being resolved (3 in this example). The top of the context stack is a direct reference to the target object of concept C.

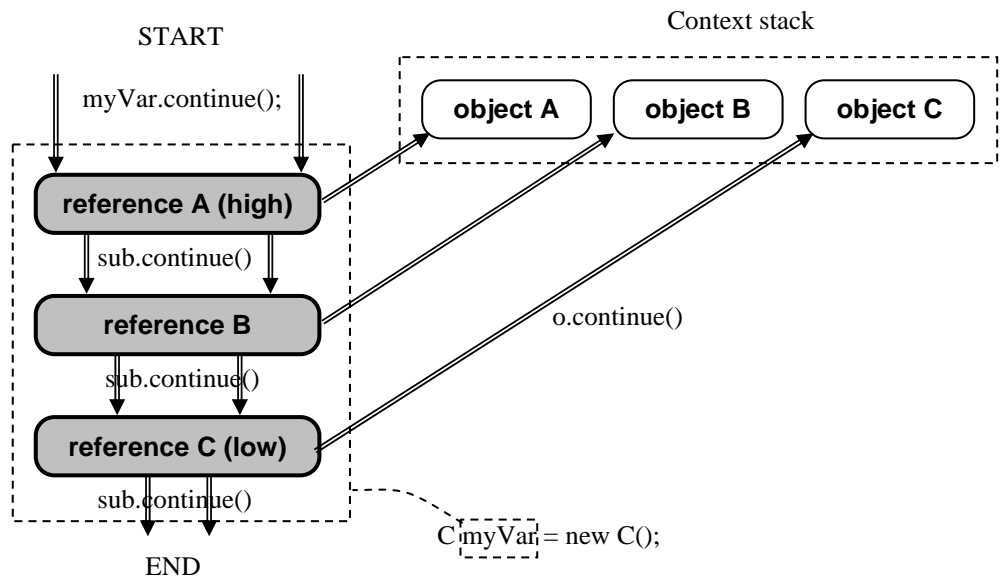

Figure 10. Complex reference resolution and context stack.

Let us consider how this sequence of resolution is implemented by continuation methods of concepts in an inclusion hierarchy. The main role of continuation method consists in restoring the primitive reference substituted by this reference. Then it simply passes control to the primitive reference and the process continues on the other side of the border in object continuation method. At this moment the resolved reference is pushed on the stack. However, if there are child objects then they also have to be resolved and hence continuation method has to call them. For example, let us assume that parent concept `Account` has one child concept `SavingsAccount` as shown in Listing 8. Continuation method of concept `Account` resolves its own account number into a primitive reference (line 7) and then passes control further (line 8). Here the mapping from account numbers to objects is stored in a global variable (line 1). Finally reference continuation method calls its child continuation method (line 9) which also has to resolve its reference segment and prepare object for access. If it is `SavingsAccount` then its continuation method resolves sub-account number (line 23) but in this case it uses its parent object for storing the mapping from sub-accounts to objects. Then as usual control is passed to the primitive reference (line 24) and finally this method gives opportunity to a possible child object to make appropriate operations (line 25).

Listing 8. Hierarchical access and context stack.

```
01 static Map map = new Map();
02 concept Account
03   reference {
04     String accNo;
05     void continue() {
06       print("=> Account: Resolve");
07       Object o = map.get(this.accNo);
08       o.continue();
09       sub.continue();
10       print("<= Account: Resolve");
11     }
12   }
13   object {
14     double balance;
15     Map map = new Map();
16   }
17
18 concept SavingsAccount in Account
19   reference {
20     String subAccNo;
21     void continue() {
22       print("=> SavingsAccount: Resolve");
23       Object o = super.map.get(this.subAccNo);
24       o.continue();
25       sub.continue();
26       print("<=SavingsAccount: Resolve");
27     }
28   }
29   object {
30     double balance;
31     Map map = new Map();
32   }
```

The most important property of this mechanism is that parent objects are directly accessible from their child objects and need not to be resolved. So each occurrence of keyword 'super' in code means *direct* access using the primitive reference from the context stack rather than new resolution via continuation method. Since normally concept functionality is based on using parent concepts, this mechanism leads to significant performance increase because it guarantees that any reference segment is resolved only once for each use of the complex reference.

## 4.4   Life-Cycle Management

To demonstrate how objects are represented and accessed we assumed that they already exist. However, before an object can be accessed it needs to be created and this moment is the starting point for its life-cycle. At the end of its life-cycle the object needs to be deleted and after this moment it cannot be accessed anymore. Thus creation and deletion are procedures which limit the life-cycle of any object in time. After creation a new reference is supposed to be valid which means that it can be used for access. And after deletion this reference is supposed to be invalid which means that it cannot be used for access anymore. One interesting detail in the life-cycle management is that creation and deletion deal with only references. That is, object existence means presence and validity of its reference (even if the object itself does not yet exist). And vice versa, object non-existence means that its reference is not valid and cannot be used for access (even if the object actually exists). This rule is a consequence of the principle that objects are not directly comprehensible and can be manipulated only via their references (see Section 2.1).

Object creation and deletion is supposed to be managed by two special methods named `create` and `delete`. These methods have the same status as another special method – the continuation method – described earlier. In contrast to the continuation methods, creation and deletion methods are normally called explicitly by the programmer. However, they possess all properties of special methods. In particular, these methods are responsible for implementation of meta-transition, i.e., they need to implement some logic of crossing the border from identity to the entity. These methods also have a special meaning for the context stack mechanism.

Listing 9. Creation method for one concept.

```
01 static Map map = new Map();
02 concept Account
03   reference {
04     String accNo;
05     void create() {
06       print("=> Account: Create reference");
07       this.accNo = getUniqueNo();
08       Object o.create(); // Go to object constructor
09       map.add(accNo, o);
10       print("<= Account: Create reference");
11     }
12     void continue() {
13       Object o = map.get(this.accNo);
14       o.continue();
15     }
16     void delete() {
17       print("=> Account: Delete reference");
18       Object o = map.get(this.accNo);
19       o.delete();// Go to object destructor
20       map.remove(accNo);
21       print("<= Account: Delete reference");
22     }
23   }
24   object {
25     double balance;
26     void create() {
27       print("-> Account: Create object");
28       balance = 0;
29       print("<- Account: Create object");
30     }
31     void delete() {
32       print("-> Account: Delete object");
33       balance = 0;
34       print("<- Account: Delete object");
35     }
36   }
37
38   Account account.create();
39
40   $ => Account: Create reference
41   $ -> Account: Create object
42   $ <- Account: Create object
43   $ <= Account: Create reference
```

Just as other methods in CoP, creation and deletion methods are dual, i.e., they have two definitions: one in the reference class and one in the object class. Reference creation method is responsible for this reference initialization which means also allocation of the primitive reference it will substitute. Reference deletion method is responsible for deletion of the resources associated with this reference including the substituted primitive references. The main role of object creation and deletion methods coincides with that of constructor and destructor in OOP. In other words, object creation method is intended to initialize the new object just after its creation and before it can be accessed (so it is the very first operation with the object). Object deletion method has to clean it up just before real deletion (so it is the very last operation with the object).

Let us assume that there is one concept `Account` (Listing 9) and we need to define its creation and deletion methods. Creation method generates a unique identifier for the new account (line 7) and then allocates system resources for this object by creating a new primitive reference (line 8). It is precisely the point where the new object is really created. Line 8 is also the point where the object creation method (constructor) is called (line 26). Object creation method simply initializes the just created object assuming that nobody has accessed it yet. The final step consists in storing the association between this account number (this identity) and the just created primitive reference it is going to substitute representing the real object. Here we simply store this pair in a global map (line 9). When this object will be accessed this map will be used to resolve object references by the continuation method (lines 13).

Deletion method follows the inverse sequence of steps. It resolves this reference (line 18) and then destroys the restored primitive reference (line 19) by executing object deletion method (destructor) just before real deletion (line 31). After deletion, this reference cannot be used anymore because the account number stored in its field cannot be resolved. Notice also that creation and deletion methods (just as continuation method) do not return any value. Instead, they are applied to a reference with the purpose to initialize or clean up its value. It is also possible define several creation/deletion methods taking some parameters.

Listing 10. Creation or reuse of available objects.

```
01  static Map map = new Map();
02  concept Account
03    reference {
04      String accNo;
05      void create(String name) {
06        this.accNo = findAccount(name);
07        Object o;
08        if(this.accNo == null) {
09          this.accNo = getUniqueNo();
10          o.create();
11          map.add(accNo, o);
12        }
13        else o.continue();
14        sub.create();
15
16      }
17      void continue() {
18        Object o = map.get(this.accNo);
19        o.continue();
20        sub.continue();
21      }
22    }
23    object {
24      double balance;
25      Map map;
26      void create() {
27        balance = 0; map = new Map();
28      }
29    }
```

In the case of concept hierarchy creation and deletion methods should propagate downward over hierarchy in the same way as it is done for continuation method. For example, if concept `Account` has some child concept, such as `SavingsAccount` then it could implement its creation method as shown in Listing 10. It is assumed that persons have main accounts with many sub-accounts. Creation method takes one parameter with the name of the owner of the new sub-account. However, it may well happen that this person already has the main account and in this case it has to be reused for creating a new sub-account. Otherwise both a new main account and a new sub-account have to be created. Thus creation procedure checks if an appropriate object can be found (line 6). In the case it is not found a new account number is generated (line 9), new object is created (line 10) and the association between them is stored (line 11). (Alternatively, we might simply call creation method without parameters as implemented in the previous example.) If the main account is found for this user then we resolve this reference by calling continuation method (line 13). Finally, we call creation method of the child reference (line 14) so that if it is a savings account then it will be created. Notice that when a child creation method is executed, its parent already exists (either new or reused) and is directly accessible because its primitive reference is in context stack.

Thus it is not necessary to really create all constituents of a complex object and each creation method can choose its own logic of creation which is appropriate for this problem domain. In particular, it is possible to implement lazy creation when we generate only a unique reference while real object creation will be performed only when this object is accessed. Another use case is where concept maintains a pool of objects as a list of primitive references. When a new object is requested to be created a primitive reference is taken from this pool rather than allocated by the system routine. Deletion method could simply mark the deleted primitive reference as unused and return it to the pool. It is also possible to assign explicitly some parent segment with valid values pointing to an existing object (such as main account). Then creation procedure will interpret it as a request to initialize only last segments. There exist also many other applications of this life-cycle management mechanism and it is especially useful for tasks where some complex logic is needed.

# 5    Operations with References

## 5.1    Reference Structure and Parameters

A length of a complex reference is the number of segments in it. Depending on the first segment type and the last segment type references may have different lengths. In order to determine the number of segments in a complex reference we will use function length(). Its argument is a variable with some complex reference and the function returns the real number of segments in it. Notice that a reference can be longer than the value derived from the type of the variable because of two factors: the first segment is a super-concept of the variable concept, and the last segment is a sub-concept of the variable concept. For example, variable savingsAccount of concept SavingsAccount may consist of a single segment and have length 1 or it may contain three segments of concepts Account, SavingsAccount and SpecialSavingsAccount and have length 3.

The length function can be also applied to a concept name and in this case it returns the number of segments from the root to this concept in the concept inclusion hierarchy. For example, length(SavingsAccount) is 2. It is also possible to provide two arguments as concept names which specify an interval of concepts in the concept inclusion hierarchy. Then this function returns the number of concepts between the first argument (excluded) and the second argument (included). For example, length(Account, SavingsAccount) is 1 because this interval includes only one concept SavingsAccount (notice again that the first argument is not included). It is obvious that

$$\text{length}(C) = \text{length}(Root, C)$$

Each reference consists of a sequence of segments where each next segment is of a sub-concept of the previous segment. It is important to have functions for determining concept for one or another segment of a reference. The next three functions, instanceof(), contextof() and conceptof() return concept name given a reference as an argument. This concept name is actually that of one of its segments.

To get concept name of a single segment we can use operator instanceof() where argument is the reference segment (it is also concept of the object represented by this reference). Let us assume that reference $a = \langle a_1, a_2, \ldots, a_n \rangle$ consists of $n$ segments. Then instanceof($a_i$) returns concept name of its $i$-th segment. If this operator is applied to the whole reference consisting of many segments then it returns concept name of the very last segment:

$$\text{instanceof}(a) = \text{instanceof}(a_n), \text{ where } a = \langle a_1, a_2, \ldots, a_n \rangle$$

In a programming language this operator returns *real* type of a reference stored in the variable. Notice that this type can differ from the declaration of this variable and may change in time because the variable may store references of different types (which are sub-concepts of its type). For example, variable account declared as having concept `Account` may contain a reference to a sub-account and then `instanceof(account)` will return `SavingsAccount`.

References do not need to start from the root segment and in this case it is assumed that some higher segments are missing, i.e., the equality $instanceof(a_1.super) = Root$ needs not to hold. In other words, a reference may include only local address relative to some context which is not stored in it. In order to get the type of context of a reference the function contextof() is used. This operator returns the type of the parent of the very first segment of this reference:

$$contextof(a) = instanceof(a_1.super) \text{ , where } a = \langle a_1, a_2, \ldots, a_n \rangle$$

If a reference includes all segments then this operator returns `Root` as its context. This means that it is a global reference with full context. It is apparent that context is always a super-concept of the real reference type:

$$context(a) > instanceof(a)$$

So it returns the type of segment which is not included in this reference. For global references it is always `Root`. In the general case the result of the operators instanceof() and contextof() depends on the real composition of their arguments.

In programming languages references are stored in variables which are declared using some type. In order to return this type function conceptof() is used: In other words, if a variable has been declared as having concept `MyConcept` then operator conceptof() will return this concept name independent of what real reference is stored in this variable. However, it will always be a super-concept of the real reference type:

$$conceptof(a) \geq instanceof(a)$$

This means that real type is a sub-concept of the declared type. On the other hand, reference context has to be a super-concept of its declared type. Hence we get the following inequality:

$$context(a) > conceptof(a) \geq instanceof(a)$$

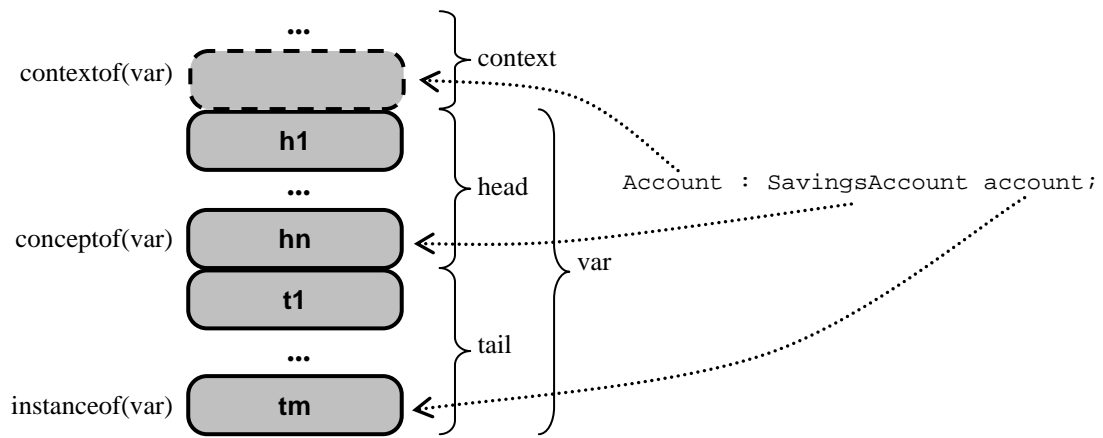

Figure 11. Reference structure and operators.

The structure of reference according to these operators is shown in Fig. 11. A reference is broken into three parts:

- <u>Context</u> is absent in the reference. The type of the last segment of the context is returned by the operator contextof().

- <u>Head</u> of the reference starts from its first segment and ends with the segment having the declared type of the variable. The type of the last segment of the head is returned by the operator conceptof().

- <u>Tail</u> of the reference depends on the real type of the reference. It consists of all segments that follow the declared type of this variable. The type of the last segment of the tail is returned by the operator instanceof().

## 5.2 Left and Right Casting

An available reference may start from one segment while we need it to have another type as the first segment. In other words, we might want to change the context of the reference either by adding new starting segments as a prefix or by removing available starting segments. This operation is called *left casting* of the reference and is written as follows:

$$\text{LeftConcept} : a$$

Here the new desirable type of context is written on the left and is separated from the variable by colon. If new segments have to be added then they are taken from the implicit context of this reference. If the reference is shortened then some starting segments are simply removed. The reference returned by this operation has the context specified in the argument:

$$\text{contextof}(\text{LeftConcept} : a) = \text{LeftConcept}$$
$$\text{where } \text{LeftConcept} \geq \text{contextof}(a) \text{ or } \text{contextof}(a) \geq \text{LeftConcept}$$

The length of the new reference depends on the new context:

$$\text{length}(\text{LeftConcept} : a) = \text{length}(\text{LeftConcept}, \text{instanceof}(a))$$

One important use of left casting consists in converting reference to a global reference by adding to it the maximal context. For example, if we have only savings account number but need to pass it as a fully qualified account description then we attach the global context as follows:

```
fullAccountRef = Root : savingsAccount;
```

Left casting can be also used to extract tail of the reference by cutting off its head part. For example, if we have an account variable and want to get only a sub-account reference stored in it (without the main account part) then it can be done as follows:

```
subAccountRef = conceptof(account) : account;
```

The operation of *right casting* changes the real type of this reference by either removing some last segments or by adding new (empty) last segments. We write the desired right segment concept after the reference separated by colon:

$$a : \text{RightConcept}$$

This operation results in a new reference with the real type equal to the specified concept name:

$$\text{instanceof}(a : \text{RightConcept}) = \text{RightConcept}$$
$$\text{where } \text{instanceof}(a) \leq \text{RightConcept} \text{ or } \text{instanceof}(a) \geq \text{RightConcept}$$

Right casting can be used to extract head of the reference by cutting off its tail. For example, if we have an account reference and need to extract only the main account identifier then it can be done as follows:

```
mainAcountRef = account : conceptof(account);
```

In this way it is also possible to get a reference to any intermediate object represented by this reference by specifying its concept.

Given two references $a = \langle a_1, a_2, \ldots, a_n \rangle$ and $b = \langle b_1, b_2, \ldots, b_m \rangle$ it is possible to find their intersection and union. Intersection of two references includes their common segments:

$$c = a \cap b \iff \forall c_k \ \exists i, j : \ \text{conceptof}(c_k) = \text{conceptof}(a_i) \text{ and } \text{conceptof}(c_k) = \text{conceptof}(b_j)$$

Union is defined dually as a new reference which contains segments of the both references:

$$c = a \cup b \iff \forall c_k \ \exists i, j : \ \text{conceptof}(c_k) = \text{conceptof}(a_i) \text{ or } \text{conceptof}(c_k) = \text{conceptof}(b_j)$$

These operations are defined only for the case where the arguments and the result contain a well formed reference, which means that the concept of each segment is a super-concept of the next segment.

$$\text{instanceof}(c_{k-1}) = \text{instanceof}(c_k).\text{super} , \ \forall k = 2, \dots$$

Assignment operation consists in copying intersection of two references into the target reference:

$$a = b \quad \Leftrightarrow \quad a_i = b_j, \ \forall i, j : \text{instanceof}(a_i) = \text{instanceof}(b_j)$$

If their intersection is empty then the target reference does not change. (Alternatively, the assignment could automatically left or right cast the target reference to get content of the source reference.) If it is necessary to copy only part of the source reference then it can be cast from left or right:

$$a = \text{LeftConcept} : b : \text{RightConcept}$$

In this case only the segments between LeftConcept (not included) and RightConcept (included) are copied (if they are present in the target reference).

Reference concatenation is equivalent to reference union where the real type of the first reference is equal to the context of the second reference. It allows us to get the first reference as the context and then attach to it segments of the second reference as an extension. Concatenation can be defined as right casting of the first reference to the type of the second and then copying the second reference to these extended segments:

$$c = a : b \quad \Leftrightarrow \quad c = a : \text{instanceof}(b) , \ c = b$$

## 5.3    Reference Length Control

In programming we do not always need fully qualified references with global context which start from the root. In many cases when the context is known or can be restored, it is enough to have only the local part of the reference. This corresponds to specifying only local address in mails which are known in advance to be used in the local context, say, in this country or in this organization. Such short references take less space and are faster during access because the context needs not to be resolved. In order to declare a short (local) reference it is necessary to specify its new context. This can be done by prefixing the variable type. For example, if the allocated variable is of type `SavingsAccount` but has to contain a reference with the context `Account` then we write it as follows:

```
Account : SavingsAccount subAccountRef;
```

This variable will as usual represent an instance of concept `SavingsAccount` but its higher segments will not be stored. By default, if the context is not specified then the variable is supposed to have the global context, i.e., it starts from the root:

```
Root : SavingsAccount fullAccountRef;
```

Given a short reference the question is how to access the represented object if some information is missing. The default rule consists in using the current context if it is absent in the reference itself. Notice that the current context is reused rather than resolved again for this reference (it is stored in context stack described in Section 4.3). This is why access on such references is faster.

The second approach to restoring context consists in specifying it explicitly when a new variable is declared. In this case we provide a concrete variable with some context instead of its type:

```
Account mainAccount = getAccount();
mainAccount : SavingsAccount subAccountRef;
```

Here new variable `subAccountRef` will be allocated using only segments after context `Account`. However, the difference is that the context itself is specified via a variable with concrete value rather than as a concept name. Thus this variable will have a concrete context rather than the current context. This context (main account) needs to be resolved when references based on it are used for access. An advantage is that one context stored in one variable can be reused by many local references.

The third approach consists in declaring some context type and then explicitly specifying context reference only when this variable is used. This approach is useful if there is one context variable and many variables storing local references. By attaching these local references to the context we can produce fully qualified references that can be used to access the objects. This is done by concatenating (extending) the context variable and the local references before the reference is used:

```
Account mainAccount = getAccount();
Account : SavingsAccount subAccount;

subAccount = getSubAccount();
mainAccount : subAccount.someMethod();

subAccount = getSubAccount();
mainAccount : subAccount.someMethod();
```

Here we declare local reference `subAccount` and get its value from method `getSubAccount`. After that we use this reference to access the represented object by attaching its segments to the context stored in variable `mainAccount`. Such an approach is more efficient than manipulating directly global references because not all segments are passed and processed and the context is attached only at the last step when the object needs to be really accessed. This approach can be even be used to concatenate several contexts to produce a fully qualified reference:

```
globalcontext : localContext : localVar.someMethod();
```

Here we store global and local contexts in different variables which are then concatenated with local object address. Notice that the operation of concatenation does not require precise match between variables – it could be made smart enough to produce a reference even if the concatenated variables overlap.

Another use of contexts consists in specifying a scope for the whole block of code rather than for only one access request. The problem here is that normally the current context is defined by the current object and all computations are executed in its context. In many cases however we need to carry out a sequence of actions in the context of another object. The legal way for doing that consists in defining a method of this object class. However, if we do not want to define a special method (for example, if this sequence of actions is used only once) or cannot define a method (if this concept is not available for editing) then the only solution consists in executing these actions in the source context. This might be very inefficient because each action requires resolution of the target reference. In order to overcome this difficulty we can change the context for the block of actions as follows:

```
MyContext myContext = getContext();
myContext : { /* block of code */ }
```

Before this block of code is entered the context `myContext` is resolved just as it is done for executing any method applied to this variable. After that the whole block has direct access to the context object. Effectively, the block of code is equivalent to a method of the concept and the only difference is that it is defined outside. In particular, all operations in the block are applied to its context by default. For example, if we need to carry out several operations within an account object then it could be implemented as follows:

```
Account account = getAccount();
account : {
  double b = balance; // = account.balance
  if(b < 100) debit(100); // = account. debit(100)
  Account : SavingsAccount subAccount;
  subAccount = getSavingsAccount();
  boolean isEmpty = :subAccount.isEmpty();
  if(isEmpty) notify();
}
```

Notice again that here the target context reference `account` is resolved only once when the block is entered and then all operations are executed using direct access. So it is not simply a shortcut since it may change the logic of the program because several operations executed individually from outside may not be equivalent to one sequence of operations executed from inside (for example, because of transactional nature of the object or security reasons).

# 6 CoP as a Generalization of OOP

## 6.1 Inheritance

In OOP inheritance is interpreted as IS-A relationship between child class and parent class. This means that if class `Ext` inherits class `Base` then we can say that any instance of `Ext` IS also an instance of class `Base`. For example, if `SavingsAccount` were defined as inheriting class `Account` then each savings account would be also an account. And if `Button` were defined as inheriting class `Panel` then again each button would be a panel. This is normally implemented as automatic inclusion of all fields and methods of the base class into its extensions. Each instance then possesses not only its own fields and methods but also those defined in its parent classes. For example, a button instance has also fields which are specific to panels just because any button IS-A panel. And any savings account IS-An account and hence each its instance has also fields which are specific to accounts.

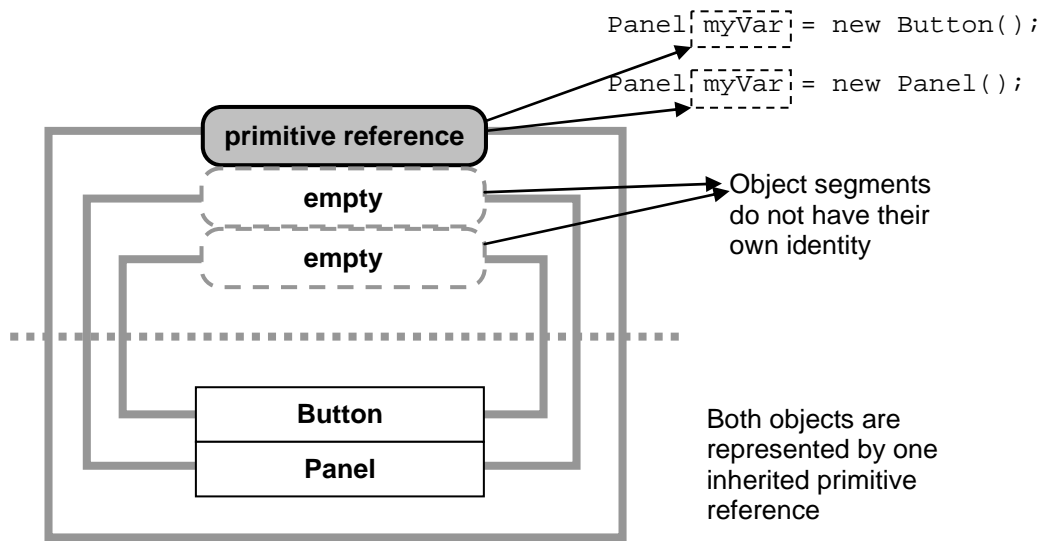

Figure 12. In OOP objects inherit base (system default) reference and exist in flat space.

One consequence of such a classical approach is that an object is viewed as indivisible entity represented by one reference even if it consists of many parts (object segments) described in different classes within its class inheritance hierarchy (Fig. 12). For each object all segments are allocated (in one interval of memory addresses) which are then considered parts of this and only this object. For example, we cannot separate panel segment (instance of class `Panel`) from button segment (instance of class `Button`) because it is one indivisible object represented by one reference. It is also not possible to share base segments among many extension segments because all segments of one object are in one-to-one relationship. For example, it is not possible to have many savings account segments for one main account segment.

Inheritance-based approach means that hierarchy exists only at class level at compile time while at run time it disappears and objects exist in flat space. Indeed, in class hierarchy many extensions may share one base class, i.e., base class is not copied or cloned for each extension. At run time the situation is different and each new extension object gets its own new base segment. It is not possible to get all extensions of one base object because the hierarchy of objects simply does not exist at run time.

Inheritance is also associated with state and functions reuse. However, an important point here is that we can really reuse object descriptions at compile time because it is enough to point to a base class. However, at run time reuse is reduced to using common behaviour (implementations of functions are not copied for each instance) while state is not reused. Instead, each object gets its own completely independent state. In other words, in OOP it is not possible to reuse (share) state of base object segments at run time.

The described classical approach to inheritance has actually deeper roots and wider use. In particular, it is used in many other disciplines like knowledge engineering (for example, frames [Min74]) or ontologies [Gru93, Gru95]. However, its central idea remains the same: fields of base classes are duplicated for each new extension. And this is precisely what is different in the concept-oriented paradigm in general and in CoP in particular, i.e., CoP changes one of the cornerstones of OOP by proposing to use inclusion instead of inheritance. An amazing thing however is that the new approach is still backward compatible with the old one, i.e., under certain simplifying conditions inclusion turns into inheritance.

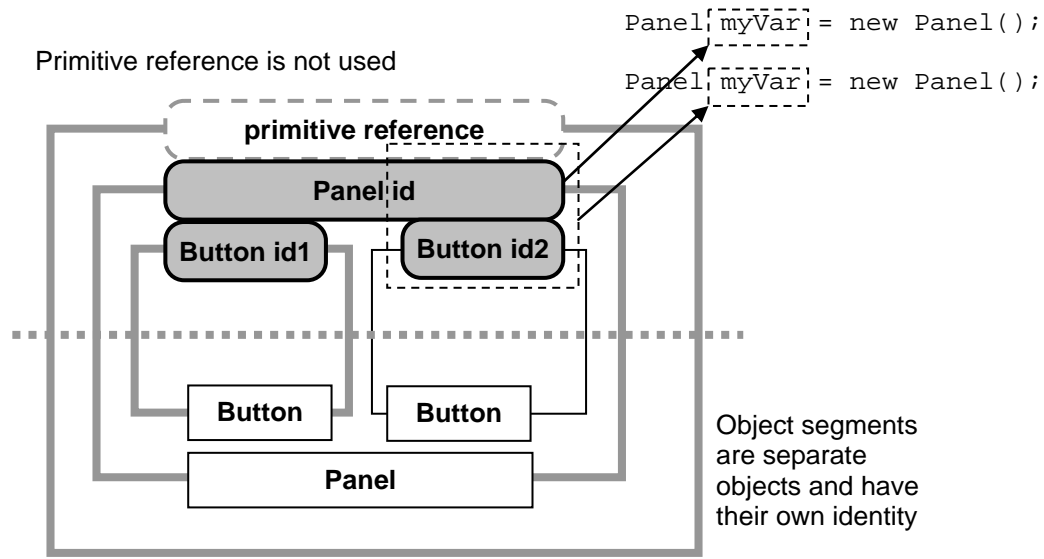

Figure 13. In CoP objects exist in a hierarchy where each segment is represented by its own reference.

The main difference of inclusion from inheritance is that it is interpreted as IS-IN relationship between elements. This means that if concept `Ext` is included in concept `Base` then we can say that any instance of `Ext` exists IN an instance of concept `Base`. Notice that we apply verb 'exists' to two instances of two concepts, i.e., they both are assumed to exist independently but in some relationship. For example, if concept `SavingsAccount` is included in concept `Account` then each savings account will exist in the context of some account. And if concept `Button` is defined as included in concept `Panel` then again each button will exist in the context of some panel (Fig. 13). The most important point here is that base objects and extensions exist independently as normal objects having their own references in the same way as cities, streets and houses are independent objects having their own identifiers. In particular, if we create a button then it does not mean that a new panel will be also created – this new button may well be created in the context of already existing panel where other controls already exist. And when we create a new savings account then the main account normally already exists and may have other sub-accounts in it.

Thus object segments in CoP exist separately at run time just as their concepts exist separately at compile time. This makes the whole picture symmetric because the hierarchy of concepts at compile time is used to model the hierarchy of objects at run time. The conception of reuse also changes its form. In CoP reuse means the possibility to use state and functions of the common context, i.e., the base object which is shared among many extensions. This significantly changes the role of the base object. Now it is interpreted as a container or environment for many internal elements. It provides useful services which can be used from inside via its object methods. Dually, it protects its internal objects from direct access by implementing some reference methods.

Inclusion can be turned into the classical inheritance if object segments in the hierarchy do not have their identity. This can be done by using empty reference classes for extensions (we still can use reference methods which will intercept incoming access requests). In this case extensions are indistinguishable in the context of their base objects and hence it is assumed that they inherit identity of the context. In other words, base object and its extensions are considered one and the same thing represented by one (base) reference. Object allocation in this situation can be optimized by putting them side-by-side in one memory interval. Obviously, here we obtain the case considered in OOP where classes by definition are not able to model identity which is provided by the system. As a consequence all object segments have the same identity in the form of primitive reference.

## 6.2 Polymorphism

Polymorphism is a mechanism which allows the programmer to manipulate objects of more specific types as if they were of the base type. In other words, we can view objects as having the base type without the need to know their concrete more specific type. For example, if we declare a variable as having type `Account` then polymorphism allows us to apply method `getBalance` to references stored in this variable even if they represent more specific account like savings account or checking account:

```
Account account;
double balance;
account = getSavingsAccount("Alexandr Savinov");
balance = account.getBalance(); // Balance of savings account
account = getCheckingAccount("Alexandr Savinov");
balance = account.getBalance(); // Balance of checking account
account = getMainAccount("Alexandr Savinov");
balance = account.getBalance(); // Balance of main account
```

In this example it is enough to know that the object is of class `Account` and that this class has method `getBalance`. However, the essence of polymorphism is that this method is not a concrete procedure but rather a placeholder or label for some general action. In other words, by applying this method at compile time we actually do not know what will happen at run time. The real action that will be executed at run time depends on the real type of the object which can be found only at run time. Thus such method calls in the case of polymorphism are intrinsically indirect. For example, if the real object is a savings account then the balance is calculated using one procedure while for checking accounts it is calculated using another procedure.

One approach to implementing polymorphic method calls consists in checking the real object type at run time and then making a decision what procedure to execute. For this purpose each object has to store information on its real type in some well known field. This technique is not very efficient and therefore in OOP normally a modified approach is used. The idea is that each object stores a pointer to a table of function pointers which are specific to its class (instead of the class identifier). All objects of one class point to one and the same table which is called a table of virtual functions or vtbl. When a method is applied to an object, the compiler makes an indirect method call using this table. Yet the result is the same – the real procedure applied to the object is not known at compile time and depends on the real object type at run time.

One property of the classical approach to polymorphism is that each new class completely overrides methods of its base class. This means that if we apply a method to an object then it is guaranteed that only one method defined in one class in the inheritance hierarchy will be executed (although the decision is made at run time). The compiler inserts a very small piece of intermediate code for each method call (actually several instructions). This code is responsible for method call indirection by choosing (dispatching) the method defined for this concrete object type.

In principle, CoP could mechanically integrate the object-oriented approach to polymorphism where child concepts completely override parent concepts. However, such a solution would be too artificial and incompatible with the main concept-oriented principles. Therefore CoP develops its own mechanism of polymorphism which generalizes the classical one. The main idea of the concept-oriented polymorphism is based on the precedence of base reference methods over child reference methods (Principle 6), which means the ability of references to intercept accesses to child references. Obviously, this means that the base reference is the very first element in access request processing chain, i.e., before the target object gets control it can somehow contribute to the processing of the method call. Thus an intrinsic feature of the method execution mechanism is that target object methods are not called directly but rather method execution is a sequence of steps. Applying a method to a reference in the source code means providing a named path for processing this request rather than specifying a concrete procedure for execution.

By intercepting all incoming access requests by the base reference, it is possible to perform some processing and then pass the request to the child reference. (When the request is being processed we can use functions of the base object.) The child reference gets the access request, processes it and then again passes to the next reference and so on till the target reference. In the simplest case parent reference does not perform any processing and simply passes it further. And only if it is the last (target) element in the hierarchy, the reference can do some meaningful actions which correspond to the type of this element. For example, let us assume that there is base class `Panel` extended by class

`Button`. If we declare a variable of the base class which is then assigned a reference to a button object then method `draw` applied to this variable will draw a button even though the variable is a panel. In OOP this is done by overriding method `draw`. In CoP, the base reference will intercept the invocation of method `draw`. Then it can check if there is an extension and decide how to proceed. In the case of the button object the base reference can simply pass this request further to the button reference segment which will be responsible for drawing this object of class `Button`. In other cases the base panel reference may do something specific to the current level before it passes request to the child element. For example, the panel might fill its background as an intermediate step.

Let us consider example in Listing 11. Concept `SavingsAccount` is included in concept `Account` (so one account may have many savings accounts as well as other types of sub-accounts). Both concepts implement method `getBalance`. The method of `Account` checks if the child object really exists (line 5) and then either returns its own balance (line 5) or the balance of the child account (line 6). In code we can declare a variable as having base type `Account` and then the balance returned by `getBalance` method depends on the real object type. If the object is of concept `Account` (line 24) then we get one behaviour. If it is of concept `SavingsAccount` (line 26) then we get another behaviour.

Notice that `SavingsAccount` assumes that there can be also internal objects (line 15), i.e., it is implemented in the concept-oriented manner where methods are intermediate processing elements getting a request from somewhere and then dispatching them to somewhere for further processing. The polymorphic behaviour is defined by the programmer who writes intermediate methods each contributing to the overall processing. We can include a new child concept in `SavingsAccount` later for example to describe some concrete savings account type and it will be incorporated into the whole access sequence by getting requests from its parent concept.

Listing 11. Polymorphism in CoP.

```
01 concept Account
02   reference {
03     String accNo;
04     double getBalance() {
05       if(sub == null) return balance;
06       else return sub.getBalance();
07     }
08   }
09   object { double balance = 10.0; }
10
11 concept SavingsAccount
12   reference {
13     String subAccNo;
14     double getBalance() {
15       if(sub == null) return balance;
16       else return sub.getBalance();
17     }
18   }
19   object { double balance = 20.0; }
20
21 Account account;
22 double balance;
23 account = findAccount(); // Real type is Account
24 balance = account.getBalance(); // = 10.0
25 account = findSavingsAccount();// Type SavingsAccount
26 balance = account.getBalance();// = 20.0
```

From this example we see that one and the same method applied to a variable of base type may cause different actions depending on the real type of reference stored in it. In CoP, such a method call is a sequence of actions associated with the reference segments. Each intermediate reference and object may contribute to the processing of the access request. As a consequence, the real composition of a complex reference influences how requests are processed. In OOP, polymorphism is much simpler and is reduced to choosing the method defined in the real object class which completely overrides its base methods. Thus the method executed by default in OOP is only the last step in a sequence of actions executed in CoP. Interestingly, CoP does not guarantee that the last method corresponding to the real object type will be reached while in OOP it is always so. The base reference methods override child methods and may finish processing at any moment without continuation. For example, the base method may raise an exception because of security constraints or insufficient resources. Or we could disable method overriding at all.

Such an approach is more flexible because request processing is distributed among all constituents at different levels rather than concentrating all the functionality in one class. An advantage of the concept-oriented polymorphism is that the programmer is able to control the whole sequence of access. However, in simple situations it is less efficient because in OOP the indirection used for calling virtual methods is optimized by the compiler.

The mechanism of polymorphism in CoP is based on the general assumption that method execution is a sequence of intermediate processing steps performed at the borders on the way to the target. A method name is a specification of the pass or tunnel through borders. Once a class implements a method with some name, all method calls having it must go through this pass. In such an approach to programming we need to describe intermediate actions executed at different borders independently of how access requests are initiated, where they start and finish. Execution of such a program is reduced to routing access requests according to their addresses stored in complex references.

## 6.3 Object Construction and Deletion

Basically object creation consists of two steps: (i) creation of object reference along with the allocation of all the necessary resource and (ii) object initialization. However, in OOP the first step is almost completely controlled by the compiler and environment while the second step is implemented by the programmer. It is quite natural separation of duties because references in OOP are not controlled by the programmer and hence their allocation is managed automatically by the system. In order to support object initialization (second procedure) classes provide a mechanism of object construction normally in the form of a special method, called constructor. What is specific in this method is that it is executed automatically just after the object is allocated. For example, in Java, bank account initialization is written as follows:

```
class Account {
  double balance;
  public Account() { // Constructor
    balance = 0;
  }
}
```

It is important to understand that constructors are not equivalent to manual initialization, i.e., having the above constructor is not one and the same as applying method setBalance(0) to the just created object. The thing is that constructor is executed at different time than any normal method. Namely, it is guaranteed that constructor is the very first method that is executed for this object after its creation. In contrast, a normal method called after object creation is executed with some delay and theoretically it may happen that some other method will intervene. So the constructor call is performed somewhere during object creation, i.e., within the system procedure, and this makes it different from any other method.

For many applications such asymmetric approach where references are allocated automatically while objects are initialized manually works perfectly. However, in many cases it is desirable or even frequently necessary to have full control over both these procedures. In other words, we would like to treat references precisely as we treat objects by having reference constructor and other attributes of the creation mechanism. In CoP this problem is solved in a principled manner by introducing concepts which consist of two classes where both reference class and object class have their own creation methods. Using the dual creation methods of concepts programmers can control not only how objects are initialized by also how they are created. In other words, it is possible to provide custom object creation procedures which substitute for the corresponding system procedures.

The same situation is with object deletion. In OOP we can control only what happens just before an object is deleted using destructor but cannot influence how concretely the object is deleted. CoP makes this mechanism symmetric by using dual deletion methods defined in reference class and object class of a concept. The reference deletion method is responsible for de-allocation of the resources while object deletion method is analogous to the conventional class destructor and cleans up the object itself.

Yet CoP life-cycle management is compatible with OOP. We can always define empty reference class for a concept and then its instances will be identified by parent references. Such concepts effectively inherit life-cycle management mechanism from the base concept. It can be rather convenient where we need a basic life-cycle management mechanism which needs to be reused for many different types of objects. It can be developed in a base concept which has a full-featured references class (and possibly does not have an object class). After that user concepts are included in it and inherit its life-cycle management functions.

# 7 Principles of the Concept-Oriented Paradigm

## 7.1 There is Always Something in-between

Informally, the classical approach to programming is analogous to the action-at-a-distance principle in classical physics dominating in 18[th] century. According to this postulate objects can interact instantaneously without any delays or interference from other objects or environment. For example, operations of reading/writing values or calling methods are considered indivisible actions. Even if we know that the elementary operations have some more or less complex implementation, we cannot use this knowledge because this layer is still closed for the programmer. Here again it is assumed that the main goal of the programmer consists in describing a sequence of operations that need to be executed. The question of how concretely these operations can be executed is already another issue. Such a classical paradigm has one big advantage: it allows the programmer to abstract from the details of implementation. Indeed, the principle of instantaneous actions is very convenient because we can concentrate on the operations that have to be executed leaving all the underlying mechanisms to someone else.

Unfortunately this advantage is reached by paying very high price. Namely, the classical approach simply cuts off all base layers so that they look like non-existing. Such an approach can be qualified as ignorance rather than abstraction. In the case of true abstraction we can develop a program as if other layers are absent but still having clear connections between layers and having a possibility to develop these layers according to the system requirements. For example, one program might have requirements to read/write operations while another program (or part of one program) might require some other behaviour for these operations. In this case we need to satisfy two conditions. On one hand we still want to work by abstracting from implementation details. On the other hand, we want to have a possibility to develop custom base layers of the system.

The classical approach works perfectly for small programs where one universal environment can serve most of the needs. However, as programs get larger it is more and more difficult to follow the principle of instantaneous actions. It is especially difficult when developing heterogeneous and distributed systems. In such systems it is already obvious that individual operations that were regarded as elementary can be quite complex and may need significant resources for execution. What is even more important is that these operations influence the behaviour of the whole system and we cannot simply ignore them. Transactions, security, persistence and many other mechanisms all belong to this category of operations which are implemented in one layer but influence the whole program. In this case one read/write operation or method call is not only rather complex action but it also serves this concrete system and has to reflect its design.

In order to overcome these difficulties the concept-oriented paradigm makes a fundamental assumption that

> *Objects cannot interact instantaneously at-a-distance and there is always some environment between them that is responsible for propagating interactions.*

This means that any operation executed in the program is not elementary and is interpreted at some other level. In particular, we cannot think of the object access process as an elementary instantaneous action. Thus assuming that instant action does not exist we postulate that the logic of propagation of interactions cannot be avoided and hence any access is by definition *indirect*. In other words, any access needs some time and intermediate processing to be performed (Fig. 14).

The fact that some basic operations are not elementary and need to be implemented at some level is not new of course. It is well known that reading/writing may need many processor instructions for execution and each processor instruction needs many processor cycles and operations at micro-level. Such a type of thinking reflects procedural programming paradigm where an operation can be implemented by means of some procedure which then can be used as one action. However, in this case we need to explicitly call this custom procedure to make use of it. In the concept-oriented paradigm this dependence is reversed. Now it is the underlying basic environment that has precedence over new layers. In other words, we are not using environment when writing a program – instead, the environment is acting implicitly behind the scenes. It takes control over the process of action propagation without any explicit instruction to do so. For example, when program methods are called the environment could intervene without any special user action. Thus procedures written in one layer are atomically used in another layer. In contrast, procedural programming assumes that user defined procedures have to be explicitly called somewhere in the program.

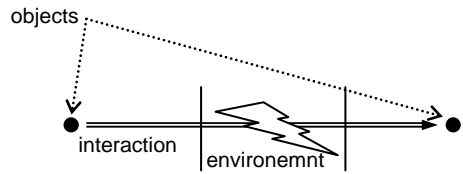

Figure 14. Interactions cannot propagate instantaneously and need some active intermediate environment.

Another characteristic feature of intermediate functions is that they have a cross-cutting nature. This means that one and the same function can be used within any piece of code. In the classical approach this phenomenon leads to numerous fragments of one and the same code distributed all over the program. For example, before an object is used it has to be locked; some methods have to be wrapped into a transaction context; an object has to be loaded from a database etc. Notice that such cross-cutting functions cannot be effectively managed using classical approach because they are not associated with any class or procedure. They are associated with arbitrary parts of the system according to some special logic which is not covered by procedures or classes (one approach to deal with them consists in using aspects [Kic97]). In the concept-oriented paradigm any code may have a cross-cutting nature if it is used as in an intermediate environment. It is a consequence of the ability of environment to actively intervene into the process of object communications. In other words, one and the same functions described in one place in the program can be automatically used in numerous places of this same program without any explicit use like method call. It is done by simply declaring this function as an environment for other functions. An advantage of using CoP in this case is that it allows the programmer to modularize cross-cutting code and put it in one place. As a result, it is not necessary to crawl through the entire program source code to find all the places where some intermediate functions are used. Instead, we move to a different model by changing the environment.

One consequence of this paradigm shift is that any explicit action written in the program entails some hidden consequences automatically triggered in the underlying environment. These consequences cannot be avoided once an environment exists. However, we can control this hidden behaviour by developing our own customized environment with the necessary functions. In any case, it is important to understand that this hidden functionality may account for a great deal of the overall system complexity. In other words, it is frequently more important what happens during access than in the end points. The process itself tends to be more important than the goal.

## 7.2 There is Always Something out There

We have postulated that objects cannot interact instantaneously and need some environment to propagate these interactions. Then the question is what this environment is and how it can be modelled. To solve this problem the concept-oriented paradigm makes the following assumption:

> *Any element exists in context of some other parent element which plays the role of environment*

Using this principle a system can be represented as hierarchical space where elements are separated by space borders (Fig. 15). In this case an interaction is only possible by intersecting the borders. And the space border is precisely the point where some processing is performed. Thus any element is not only a starting/ending point for interactions but also a border between elements which mediates interactions among them. The shift of paradigm is that now the system functionality is concentrated on space borders rather than in objects themselves. And this functionality is automatically triggered whenever any process intersects it. In contrast, in the traditional OOP approach the functionality is supposed to be concentrated in objects (end points) and it is activated only explicitly by calling methods or sending messages.

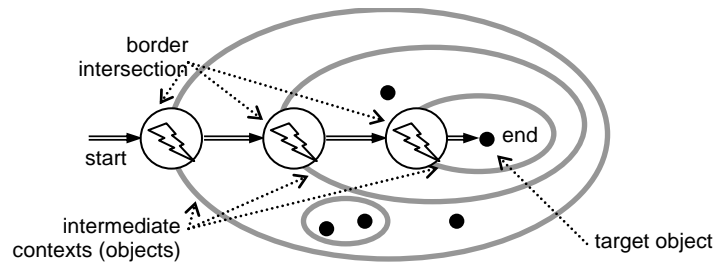

Figure 15. Computation is a process of border intersection during object access.

Parent elements play a role of environment for their child elements and hence they are responsible for transfer of interactions between them. If one child element needs to interact with the other child element then they use their parent element which is actually the only member of the system that knows how to connect them. During the process of interaction transfer between its children, the parent element can actively participate in processing the requests. On the other hand, context can provide useful services to its child elements which means that environment not only implicitly intervenes into the process of information transfer but also helps its children by providing basic functions that can be explicitly called from inside. The behaviour of any element in such a hierarchical environment is always *context-dependent*, i.e., one and the same element can behave differently in different contexts and depending on the space geometry. The context is analogous to physical laws or laws in a state: we cannot change it if it is already available but we can adapt to it and use it for our goals. Another informal analogy is relativity theory which says that objects in space change its geometry and the geometry influences objects that exist in it. In CoP the situation is the same: objects are not simply end points but rather implement functions of space with some specific geometry and influencing other objects.

Another consequence of having the hierarchical structure of elements is the change of extension/specialization principle. In the concept-oriented paradigm extension/specialization means placing an element into a context or putting into a parent environment. In particular, this allows for sharing base elements.

Traditional approach to programming consists in describing a sequence of actions which can be defined as procedures or class methods. A procedure or method call in this case results in an immediate execution of its operations. Although in reality it is not so and there is always something out there, the existing approaches simply ignore the events happening under the hood. CoP significantly changes this view and here again we see the paradigm shift. Any operation is actually a *sequence* of actions executed on the way from the current location to the destination. The programmer specifies only the final goal (object) and the type of processing (method) while the real processing depends on the position of the target object in the hierarchical space.

In particular, the programmer cannot say precisely what happens after a method is applied to an object. A method call is a final goal and its execution depends of the intermediate elements of the system, i.e., what borders will be crossed. For example, crediting a bank account involves one arithmetic operation applied the account object. In OOP calling this method would result in precisely what is written in its definition, i.e., executing this arithmetic operation. In CoP this arithmetic operation will be only the last action in a sequence of possibly hundreds or thousands intermediate operations executed during object access. All these operations are executed transparently behind the scenes but they are written by the programmer. In such an approach behaviour of the program is less deterministic. We cannot say where one or another function will be executed because each intermediate element will make its own contribution. The programmer describes only local behaviour of intermediate elements and these functions then are activated automatically during object access.

## 7.3 Everything Has an Identity and Entity

We have postulated that objects exist in a hierarchical space where they represent borders actively intervening in the process of interactions between objects. However, to interact in a hierarchical space, objects need unique addresses which could represent them in any part of the space. Without such an address system interactions are not possible because it is not possible to specify target objects. Thus having an address is one of the necessary requirements to any element. And vice versa, if there is an address then it has to represent some object. In other words, if we define an object, then we have to think about its identity and once we define an identity we have to think about the entity it represents.

The shift of paradigm is that we think in pairs rather than using separately identities and entities. A single reference without an object and a single object without a reference are degenerated constructs. In the concept-oriented paradigm they are two parts of one and the same thing and should be modelled together. Similarly to the principle "everything is an object" in OOP, we use the principle "everything is a pair of one reference and one object" in CoP.

One of the main contributions of the concept-oriented paradigm is that it completely legalizes references as a crucial element of any system (while CoP proposes to use concepts for their modelling in a program). References are made first class citizens with the same rights as objects. The shift of paradigm is that having only objects is not enough to efficiently describe behaviour of a system. Such an object-oriented picture is not complete what is especially problematic in complex systems. It can be completed by adding the notion of reference as an object representative. However, the main contribution of the concept-oriented approach here is that references are not simply recognized as an important element but made dual to objects, i.e., they may possess behaviour just as objects. In some sense references can be viewed even more important elements than objects because a system can exist without objects but it cannot exist without references. Moreover, a system consisting of only references may possess rather complex functionality. Another important property is that development of a system starts from developing the structure and functions of its references, i.e., it is more important how objects are represented and accessed while object functions can be defined later.

The duality of references and objects can be considered a continuation of a very general and deep principle of Separation of Concerns formulated by Dijkstra [Dij76]. The main idea of this principle is that any problem or system functionality can be viewed from different points of views or concerns. One specific feature of the concept-oriented paradigm is that we distinguish *two* orthogonal concerns any program consists of: behaviour of references and behaviour of objects. To develop a program functions of references are as important as functions of objects. In this sense it is very important to have adequate and convenient means for modelling inseparable unity of these two cross-cutting concerns and concepts in CoP satisfy most of the necessary requirements. The duality of references and objects creates a nice yin-yang style of balance and symmetry between two sides of one reality.

An important feature of the concept-oriented paradigm is that it proposes to use one construct, called concept, for modelling both concerns and both sides of any element. Informally, the relationship between concepts and classes is analogous to that between complex numbers and real numbers in mathematics. Like complex numbers consisting of imaginable and real parts, concepts involve two constituents but are manipulated as one construct which may exhibit properties of both references and objects. In the same way as complex numbers are much more expressive and natural for many difficult mathematical tasks, concepts are much more expressive and natural in computer science including programming, data modelling, system analysis and design.

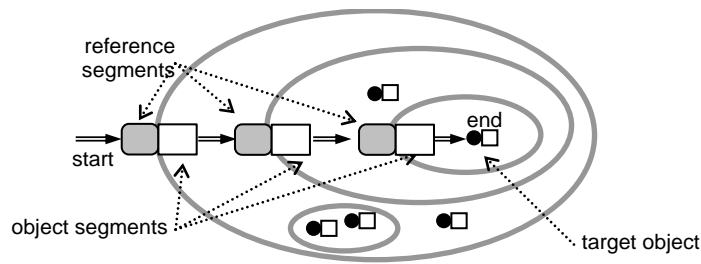

Figure 16. Objects in space are identified by their hierarchical addresses

Taking into account that the space where objects exist has a hierarchical form, references are also arranged hierarchically, i.e., any reference has a parent reference (Fig. 16). Thus it is important to understand that any reference is a *relative* representation of an object with respect to some context. This relativity cannot be avoided and even the root of the hierarchy has a restricted scope. Thus by choosing some root we simply outline the scope of the whole system or model by postulating that everything in this system will happen within this context.

A consequence of having references and objects is the duality of methods. This means that in the general case all methods have two definitions: one in the reference class and one in the object class. Reference methods are used to enter the context while object methods are used as services provided by this context to internal elements. In contrast, in OOP there is only one type of methods.

## 7.4 Everything is Relative and Virtual

The concept-oriented paradigm uses inclusion relation to describe hierarchical structure of space where inclusion generalizes inheritance. However, inclusion relation is only one of two structuring relations in this approach and the second one is substitution relation which does not have classical analogues. Substitution is a completely new relation that must be taken into account in any system design. Thus the concept-oriented paradigm uses two relations: inclusion and substitution. The main role of inclusion consists in extending the existing coordinates by introducing internal coordinates which are concatenated to the parent as additional segments. The main role of substation consists in introducing new levels of indirection elements of which are used instead of the elements of the substituted level. In other words, inclusion adds new levels while substitution replaces existing levels. New levels produced by inclusion relation are used *in addition to* the existing parent levels while new levels produced by substitution are used *instead of* the existing parent levels.

Substitution relation introduces a new dimension into system analysis and design because it allows the developer to describe *virtual spaces*, i.e., substitution is a mechanism of virtualization. For example, *instead of* using physical memory address we can introduce our own references which substitute them. And then *instead of* using these custom references we can introduce names for identifying objects. Each new substitution level is a new virtual address system. It is called virtual because it does not exist in reality, i.e., it is our own internal convention (a convention within some scope or context). Such virtual addresses do not have any indication what kind of reality they are based on. The shift of paradigm here is that developing such virtual spaces is one of the main concerns in system design. In other words, we say that any concrete system needs its own virtual space where its elements will exist. We do not develop a system in terms of computers, memory, disks and other physical entities. Rather, we create our own virtual environment which reflects the problem domain. And after that this virtual environment can be bound in one or another way to the physical environment(s) or a base virtual environment.

Program development in OOP focuses on describing object behaviour via classes which can be expressed as follows: "describe your classes and you will get your system". In contrast, in the concept-oriented paradigm the focus shifts in the direction of virtual address spaces and one of the main concerns consists in describing their layered structure: "describe your virtual address spaces and you will get your system". Defining the virtual address system is what the concept-oriented development process should start from because objects cannot live in vacuum and need identifiers. They cannot be homeless and need some environment providing all the necessary facilities like life-cycle management. The pair of inclusion and substitution relations provides the mechanism for designing such object homes with virtual addresses and other services. Notice that the objects themselves can be quite simple and most of the complexity is concentrated in the virtual environment and this is why most of the attention should be paid to its development. Here we see a striking similarity of CoP with an old idea expressed by Butler Lampson (who attributes this saying to David Wheeler) in one of his aphorisms: "All problems in computer science can be solved by another level of indirection". A distinguishing feature of CoP is that it proposes a concrete solution for dealing with indirection, particularly, by bringing substitution relation into the theory and practice of programming.

# 8 Related Work

## 8.1 Hardware Level

Access indirection exists at all levels of the system organization including physical interactions and even deeper just because instant access is an abstraction and any access requires some intermediate environment and some time to propagate. However, here we start the discussion from the hardware level of the system organization ignoring all the underlying levels such as implementation of digital micro-circuits.

It is frequently assumed that memory addresses provide direct access to its contents. However, it is an abstraction which is valid only from the point of view of the system programmer. For the processor each such access means quite a lot of work. First of all it is important to understand that (in contemporary architectures) each address is actually a location in a *virtual* address space, i.e., it is an abstract space which is not directly bound to the real memory. In 32-bit architecture one address is a 32-bit word (4 bytes) which describes the location of one byte in memory. However, each application has its own virtual address space, i.e., they exist in different contexts. An application can use such virtual address to manipulate its contents by reading/writing it. However, all these manipulations are made in a virtual address space rather than in physical address space. This means that the application does not actually know where the necessary cell is located in physical memory. So a virtual address *substitutes* a physical address.

As we already noticed, having a virtual address system has two consequences. First, we are not bound to the physical reality and have more freedom in manipulating data. Second, each access is more complex and less efficient and it is the task of the processor to resolve virtual addresses into the real physical location. In order to support two spaces and indirect access, processors provide several mechanisms. First of all, it is necessary to understand that the context for each virtual address space needs to be somehow expressed. It is normally a kind of starting address of the virtual address space while virtual addresses are offsets applied to this origin. This starting address is manipulated by the processor with special commands normally in kernel mode using the operating system kernel routines. For example, when an application is loaded, the kernel allocates a virtual address interval in the physical memory by initializing the corresponding processor registers. After that the processor can automatically translate virtual addresses into physical addresses in user mode.

Resolution of virtual addresses is implemented at hardware level using the values of the processor registers and other data structures. In other words, processor knows that the addresses in user registers need to be resolved into physical addresses and does it transparently for each access. It is important also that processor carries out many other useful operations for each indirect access. In particular, it checks security conditions, i.e., it checks if this application is permitted to access the addressed cell using the specified operation. It also checks if the physical memory is really available. If not, then it raises an exception which is caught by the operating system. This mechanism is used to implement swapping where pages of memory are stored on disk and the memory is discarded or used for other applications. This allows having small physical memory which serves large virtual address space. For example, 64-bit virtual addresses allow the application to address memory intervals having size $2^{64}$. Yet the real physical memory is normally much less.

Virtual address may have some structure which needs to be mapped to the physical address structure. For example, earlier 16-bit processors manipulated addresses consisting of one 16-bit segment and one 16-bit offset stored in two registers. These two elements of the address produced 20-bit full address by means of 4-bit shift executed by the processor during access.

Thus it is important to understand that the conception of indirect access, custom references and address space virtualization already exists at the hardware level. However, its specific feature is that it is already hard-coded into the processor architecture and cannot be changed. Indeed, it is not possible to introduce a custom virtual address space or a custom security mechanism for memory access. The main goal of our research consists in providing means for extending existing representation and access mechanisms.

## 8.2    Software Level

Custom mechanisms for indirection representation and access can be created at different software levels starting from operating system and ending with custom libraries. In this case various computer resources are represented by special identifiers which are not directly connected with their internal representation. Such identifiers are supposed to be used externally while internally they substitute a direct identifier and there is some mechanism that is responsible for their resolution. This approach makes representation and access even more indirect with respect to hardware level and hence it is more abstract/virtual and less efficient.

Operating systems normally provide their own mechanisms for memory access. One of the most wide spread technique for memory management is global heap. The idea is that operating system provides the available memory for use by applications. Each element in the heap has its own unique identifier, called memory handle, which is however not directly connected with the corresponding address in memory. Memory handle is an identifier in a virtual address space created and managed by operating system. In order to access a piece of memory represented by such a handle it has to be locked and resolved by the application. Operating system keeps a mapping from the space of memory handles into the memory addresses.

In addition, operating system might also provide other types of containers with their own address spaces. For example, for small objects an application might use local heap. Another type of virtual memory is built for file access. In this case operating system provides the space of physical disk block for use by applications. However, it is done with the help of any address space where intervals of disk blocks are allocated as files. Each file is uniquely identified in this virtual space by its file handle. File handle are not directly connected with file blocks on disk, i.e., each time we need to access a file, the file system resolves its handle to the real location on disk.

Custom mechanism for access indirection can be introduced at the level of middleware. The idea of middleware-based approaches consists in creating special software environment where a conventional program will run. In particular, such an environment offers a number of functions that are intended to support indirect representation and access. This special environment can exist and be accessible to running programs in very different forms, for example, as part of an operating system, an object container, a service, a dynamically or statically linked library etc. One wide-spread class of middleware is techniques for remote procedure calls. Examples of such middleware platforms are CORBA and RMI/EJB [Mon06]. These environments provide facilities for creating remote references and then making transparent method calls. As a consequence access to objects is even more indirect. Such middleware platforms may fit well to the purposes of one system but may be inappropriate for another system. In other words, it is yet another standard level of indirection which however exists separately from the program and hence cannot be easily adapted to the purposes of each concrete program.

Methods of access indirection can be made part of a programming language infrastructure. An advantage is that the language run-time environment is closer to the programmer and hence it can be better controlled in comparison with the middleware-based approaches. One technology that can be used at this level consists in changing the behavior of a language from this very language. This allows the programmer to adapt this language features to the needs of each concrete program. These possibilities are provided by reflective environments and metaobject protocol [Kic91, Kic93]. Normally programming languages are defined in such a way that their behavior cannot be changed. In particular, we cannot change how objects are represented and accessed because it is hard-coded into the language and its environment. The reflective approach allows the programmer to change this environment and to change the way how the language constructs are interpreted.

Another wide-spread approach to access indirection that belongs to this category consists in providing the mechanism of access interception at the level of the language run-time environment. In Java Virtual Machine this mechanism is called *dynamic proxy* [Blo00]. It exists also in C# where its functionality is implemented in the `RealProxy` class.

## 8.3 Design Patterns

Above we described non-language approaches which can be quite useful if indirect access is an auxiliary technique. However, if indirection is regarded as one of the primary mechanisms then it should be supported at the level of the programming language itself. This allows the programmer to express an *arbitrary* logic of indirection without restrictions imposed by the hardware, operating system, middleware or a library.

The simplest approach to using a new mechanism in a programming language consists in following some discipline or pattern [Gam95]. One such wide-spread pattern for implementing indirect access in an object-oriented programming language is called *proxy*. Proxy is a special class that emulates the interface of the corresponding target class but inserts some intermediate functionality. (Proxy as a pattern should be distinguished from dynamic proxies as a built-in feature of run-time environment.) These intermediate functions of the proxy class are called before the target methods and hence they effectively intercept all target object method invocations. For example, if we need to indirect access to class `Account` we could define its proxy class `AccountProxy` implementing the same methods. The trick here consists in using proxy class instead of the target class, for example, we need to explicitly use class `AccountProxy` instead of class `Account`. Thus it is not a real interception but rather a normal sequence of method calls using static and explicit substitution. In other words, in the source context a reference to the proxy instance is created and hence the methods of the proxy are called when it is used. Then it is the task of the proxy to decide what to do if some its method has been called. Normally, after some processing the corresponding target method is called. One disadvantage of this approach is that it requires significant manual support and is not very general. It is more a special technique or specific programming pattern rather than a programming paradigm. Here are other disadvantages of this approach:

- If the target class changes then its proxies need to be updated because this pattern is a manual implementation of a discipline while transparency needs to be supported by a mechanism of the programming language.

- One limitation is that a proxy is developed for one target class because it is aware of and explicitly simulates its behaviour. A true interceptor has to be able to intercept method calls to *many* different classes.

- It is difficult to impose behaviour in a nested manner (creating a proxy for a proxy) because it requires even more manual support and such a program is even more sensitive to changes which have to be propagated over the source code.

- This approach allows the programmer to implement indirect access but it does not provide means for modelling references which are passed and stored by value instead of native references. A reference to a proxy is still a normal native reference.

Dependency injection (DI) and inversion of control (IoC) are patterns which are similar to context dependence in CoP. These patterns can be used to achieve loose coupling of components by injecting some resource or capability. In this case it is the architecture that links the components rather than the components themselves have to link themselves. In this sense it is also similar to plug-in architecture where components are linked via the available framework. In CoP, the architecture and services are provided by base concepts while child concepts play the role of components. In this case base concepts are responsible for many functional aspects of their children.

## 8.4 Language Level

The main goal of the research described in this paper consists in developing *language* means for describing indirect ORA function. An advantage of having language support is that the programmer or modeller can describe *any* desirable configuration of indirection which is required by the system without any restrictions from the available hardware/software/middleware environment. In this section we describe existing language-based techniques that can be used for this purpose.

One well known technique which is intended to support indirect representation and access is referred to as *smart pointers* [Str91]. In C++ smart pointers are based on using the mechanism of templates (parameterized classes). The motivation for smart pointers is very similar to our motivation and consists in the desire to have a control over the process of object access which is absent in the case of native pointers. At the same time, it is desirable to have an illusion of direct access, i.e., smart pointers have to be analogous to native pointers when they are used for object access. In other words, we need to be able to store and pass smart pointers as if they were native pointers although they store some other data or functions.

Smart pointer is an instance of a class that is developed to model references which are passed by value and provide access to an object. For example, we might define class `MyReference` and then instantiate it whenever we need to represent an instance of some target object. Here we already assume that such a class will serve many different target object classes by encapsulating general behaviour common to all of them. However, although the reference class has a generic form, this solution assumes that we still need to know the target class for the smart pointer to work properly. In other words, smart pointer classes are normal classes which are defined independently of the object classes they will represent but these target classes must be known at some moment. For this purpose the smart pointer class is parameterized by the name of the target object class using the mechanism of templates. This parameter is then used in implementing methods of the smart pointer class. When a new smart pointer has to be created we must provide the name of the target object class as the value of the parameter, for example, as follows:

```
MyReference<Account> account(new Account());
MyReference<Person> person(new Person());
```

Thus in this approach we must specify both the target class and the reference class for each new instantiation, i.e., references class and object class are paired at the time of use for each individual variable. Another property is that an object could be represented by many different reference types and one reference type can be used for many different object types. The real interception is performed by overloading the universal access operator (dot or arrow in C++). This means that each time a method is applied to a smart pointer this overloaded operator is called and then executes the intermediate actions. Normally it resolves the reference by finding the location of the target object and then calls the target method.

The same idea of having a reference class parameterized by a target object class is implemented in the Transframe programming language [Sha]. The difference is that in Transframe, *referentail classes* have more support by the language, i.e., it is a built-in feature while in C++ it is pattern based on a more general mechanism of templates. In this language we can mark a class as intended to represent other objects as follows:

```
class MyReference is referential {
  private:
    obj: ObjType;
  public:
    enter(path:char[]) { ... }
}
```

Notice that the referential class is still defined independently of any target object classes, i.e., it is a normal class with fields and methods. An just as for C++ smart pointers, when a new instance of an object has to be created we need to specify both the reference class and the target object class:

```
account: MyReference of Account;
person: MyReference of Person;
```

The difference from smart pointers is that the parameterization is supported by the programming language rather than uses the general mechanism of templates. Below we list main properties of these two approaches and their difference from CoP.

- Smart pointers and reference classes in Transframe use independently declared reference classes and object classes while in CoP these two classes are defined as parts of one more general construct (concept).

- Smart pointers and Transframe assume that a new instance of an object requires two parameters: a reference class name and the object class name. Thus association between these two classes is established at the moment of creation. In CoP the association between the reference class and the object class is established within the concept at the moment of its declaration while creation is performed precisely as in OOP using only the target object class.

- Smart pointers and references in Transframe can be viewed as explicit proxies because they are created as objects of certain class which are then used instead of the target objects. In CoP we have an illusion of working directly with the target object while the indirection is completely hidden.

- Smart pointers and Transframe allow the programmer to implement simple substitutes passed by value but it is difficult to implement hierarchical (complex) references consisting of several segments. And it is even more difficult to implement objects consisting of several separate segments.

- Smart pointers intercept method calls by overloading access operator. In CoP the interception of individual methods is performed by reference methods of the concept which have precedence over object methods and have an opposite overriding direction. General interception can be also implemented via continuation method.

There exist also other language-based approaches to automating indirect access. For example, one general purpose method, called attribute-oriented programming, consists in using annotations or tags to mark up parts of the source code where additional intervention is needed. In particular, this allows us to make use of custom (external) source code processing units responsible for reference creation and interception of access.

Another interesting approach, called language-oriented programming (LOP), consists in developing custom languages for each task or problem domain [War94, Dmi05]. In LOP the programmer could develop or use libraries of languages just as libraries of procedures or classes are used in procedural programming or object-oriented programming. In particular, such a custom language could support functions of indirect representation and access. However, here we still have a problem of creating such custom languages.

The next general approaches that could be used to implement different elements of the mechanism of indirect access consists in using *mixins* (abstract sub-classes) [Bra90, Sma98], subject-oriented programming [Sop] and multidimensional separation of concerns [Mdsc]. These methods allow the programmer to describe how behavioural granules have to be distributed throughout the system. However, they are not targeted at the problem of indirection of representation and access and provide only more or less convenient means for implementing different representation and access patterns.

One of the most interesting recent approaches to programming is aspect-oriented programming (AOP) [Kic97]. Aspects describe intermediate functionality (and data) injected into the points in the program which are specified by means of regular expressions. Thus aspect can be viewed as a special programming construct that modularize intermediate functionality. An important property of this approach is that aspects know explicitly the points where the intermediate functions will be injected while the target classes do not know what other code will modify their behaviour (Fig. 17). Such a structure of dependencies between the module with the code to be injected and the modules where it has to be injected can be viewed as declaring all the target classes within the aspect (the target classes being unaware of this aspect). In this sense CoP is characterized by the opposite direction of the dependence. Namely, the module with the code to be injected (parent concept) is unaware of the points where it will be used (child concepts). What is similar between AOP and CoP is that method invocation is indirect and can trigger quite complex intermediate actions. However, the principles behind this indirection are quite different.

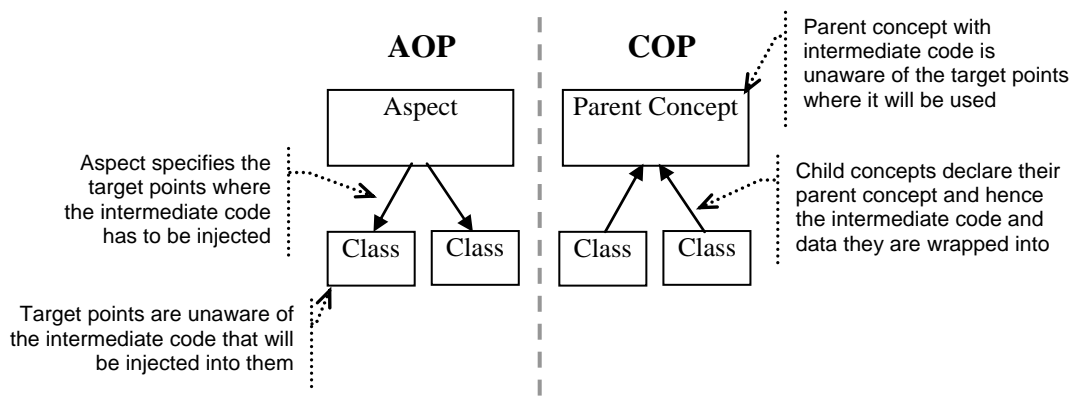

Figure 17. Aspect-oriented programming vs. concept-oriented programming.

The mechanism of dual methods in CoP is similar to *super/inner* methods of classes [Gol04] which is implemented in the Beta programming language [Kri83, Kri87, Kri89]. In particular, the inner methods are designed in such a way that they implement the same sequence of access as that in reference methods. However, the mechanism of super/inner methods is implemented as an addition to normal classes. Hence it can be viewed as an enhancement to OOP aimed at providing means for object protection from outside. In CoP, this behaviour is implemented using a completely different approach, namely, by means of concepts.

The concept-oriented approach relates also to so called *context-oriented* methods which are aimed at bringing context dependence into programming [Cos05, Gas98, Rak02, Kea03]. These methods introduce languages constructs and mechanisms which allow the programmer to put objects in a context changing their behaviour at run-time. For example, in the ContextL programming language it is done by means of the keyword 'in-layer' while in CoP we use 'in' which generalizes inheritance and 'super' to access the context. The context-orientation also relates to a technology known as dependency injection.

There exist also approaches to programming having the same name but which are actually based on very different notions and do not relate to our work, for example, [Voi92] and [Mcc99]. The mechanism of concepts exists also in generic programming where it is used to describe a set of operations supported by a type [Gre06].

If we use only references in CoP then we get an approach where only values are manipulated, i.e., any operation is applied to only one reference while references themselves are copied by value. Since objects are not in this approach, references are interpreted as data passed by value. In this sense it is similar to functional programming.

## 8.5 Conceptual Approaches

The question what is an object and what is its identity is of crucial importance in computer science. Although these questions are of approximately equal importance, the first question about the properties of objects has been paid much more attention and this direction has much more theoretical and practical results. In particular, two major paradigms – object-orientation and relational model – are almost completely devoted to the issues concerning object/entity modelling and do not provide similar means for identity modelling. Although general means for identity and access modelling have not been developed, this problem was extensively studied from philosophical and conceptual points of view.

Very interesting model where objects and their identity are connected within one model was proposed by Kent [Ken91]. In this model scope is analogous to what we mean by space or context, i.e., it is assumed that everything exists in some scope. However, this scope is not a normal element – it is an additional construct in the model. In contrast, in the concept-oriented model context is just normal element, i.e., any element exists in some context and any context is an element. Scope in Kent's model has the same role with respect to references as in our model: "A token belongs to a scope, which determines its status and meaning as a reference, based on the status and meaning of its corresponding symbol in that scope." However, this role is assigned in an informal manner while in our approach it is a concrete mechanism implemented using special functions and obeying a concrete sequence of access.

Another feature of the Kent's model is that scopes are not intrinsically nested. Strictly speaking, it is noticed that scopes can be nested: "Scopes could be nested, but that's beyond the concern of this paper." Yet the nesting is not used a high level principle while in the concept-oriented model this relation is of primary importance, i.e., we assume that elements cannot exist without a hierarchy and without a parent element (context).

The next property of the Kent's model is that it does not have substitution relation among elements, i.e., objects have identities but it is not described how these identities are used for access. This means that identities have a mechanism of access as their intrinsic or primitive property. In our approach substitution relation is of the corner stones. It should noticed however that Kent's model allows for synonyms.

Another interesting model is proposed in [Wie95]. The authors provide a definition of an object identification schema, study its properties and compare with the existing identification mechanisms such as oids, keys and surrogates. In great extent this paper consolidates what was known at that time about object identification. However, it does not integrate the theoretical conceptual analysis of the role, applicability and limits of object identification into any practical framework or setting like programming approach or data model.

This model uses the following elements:

- Value space *V* consists of abstract entities called values. They are supposed to be unobservable and unchanging. For example, number 2 as a value is distinguished from its representations that may have different forms.

- Symbol set *S* consists of symbols where each symbol is a type that may have many instances, called symbol occurrences.

- Symbol occurrence is supposed to be an observable part of the world. For example,"E" is an occurrence of symbol 'E' representing character E as a value.

- An object set *O* consists of all possible objects.

In this approach naming schemes connect values and objects using some relation. A particular case of naming scheme is an oid schemes.

## 8.6    Concept Related Approaches

There can be two approaches to programming based on using concepts defined as a pair of one reference class and one object class which are referred to as CoP-I and CoP-II. These approaches depend on the role played by these classes and their responsibility.

CoP-I [Sav05] assumes that references of a concept represent objects of its *child* concepts (not this concept). For example, if we define concept `Account` in CoP-I then its references are intended to represent child objects such `SavingsAccount` or `CheckingAccount`. How then objects of this concept are represented? They are represented by references of the parent concept. For example, account objects could be represented by references of concept `Bank`. Thus concepts in CoP-I describe one space object with a set of internal references for representing internal objects. This space object knows references of its internal objects but does not know what kind of objects they will represent. What is important is that an object is responsible for managing a *set* of references.

CoP-II [Sav07, Sav08] assumes that references of a concept represent objects of this same concept. Thus concepts in CoP-II describe one element which consists of one object and one reference. If we represent elements as a hierarchy then the difference between these two approaches can be represented as shifting of concept vertically (Fig. 18) so that its reference class and object class correspond to different parts of this hierarchy.

Both approaches have some advantages and disadvantages. CoP-I has some very attractive features and on earlier stages of development we thought that the first approach is more perspective. Later on, when trying to remove some subtle problems we gradually switched to the second approach described in this paper.

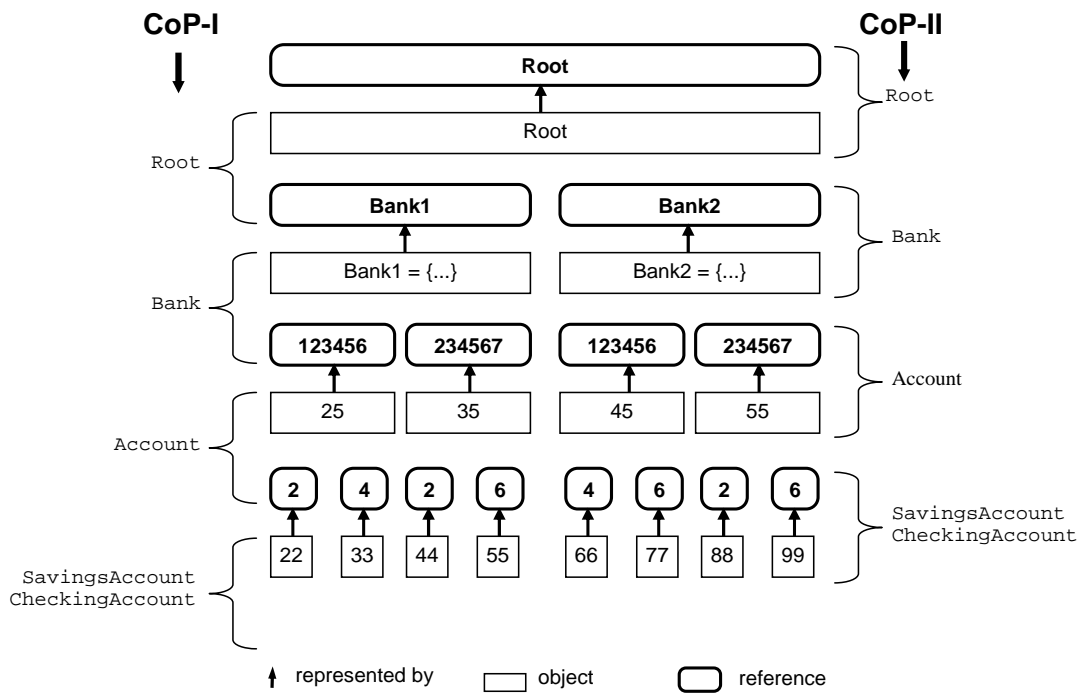

Figure 18. Two approaches to concept composition: CoP-I (left) and CoP-II (right).

There exist other concept-related approaches which use term concept for their constructs, mechanisms or general ideas. However, these approaches are not related to our method. For example, B. McConnell proposes an approach to programming which is also called concept-oriented programming [Mcc99]. This method uses DNS-like system which extends object-oriented programming languages. This system stores reusable code that can be loaded from programs as a library. An advantage is that programmers can write tiny programs that load these libraries.

Other concept-related approaches include concept-oriented databases, concept-oriented logic programming [Voi92], concept programming and XL programming language.

# 9    Conclusions

In the paper we proposed an approach to programming based on using concepts where a concept consists of two classes: a reference class and an object class. Both the constituents of concepts have their own members defining their individual structure and behaviour. What is important is that concept as a programming construct provides adequate means for modelling pairs of objects and references connected by inclusion and substitution relations.

One of the main achievements of the proposed approach to programming is that references are completely legalized and made first-class citizens of the program along with objects. We completely abandon the idea of having entities (objects) and identities (references) as separate things and follow the principle that any element is a pair of an identity and an entity.

An amazing feature of concepts is that they smoothly generalize classes and hence the concept-orientation can be viewed as a generalization of the object-orientation which has been a dominant programming paradigm for many decades. This generalization is not restricted by the simple fact that concept can be obtained from class by attaching a reference class to it. It covers also such issues as inheritance and polymorphism which take new more general forms within CoP.

Although in the paper we have focused manly on technical issues, this approach is not simply an additional mechanism or technique that can be implemented in existing programming languages. Rather, it has significant influence on the way how a complex software system is viewed and a computer program is being developed. Such a change is normally called a paradigm shift. In fact, this change of paradigm preceded the development of the described language means, i.e., first, we formulated general principles and then developed several versions of language mechanisms to support them. One of such changes is that programming in CoP is targeted at describing virtual address spaces for objects. Thus abstraction is reached by introducing new indirection levels using custom references and access procedures of the virtual address spaces. In contrast, existing approaches provide abstraction via new object classes and procedures.

Another important notice about the proposed approach is that this research has been performed in close connection with the corresponding data modelling techniques because they share the main principles. Such an approach to data modelling is referred to as concept-oriented data model (CoM) [Sav05a, Sav05b, Sav06, Sav07a]. Thus concept-orientation is not restricted by programming languages and plays very important role in data modelling. The concept-oriented data model consists of two parts: logical and physical. Physical part is almost precisely what is described in this paper while logical part covers mechanisms which are specific to data modelling only such as data semantics and operations with data.